\documentclass[11pt,a4paper]{article}
\pdfoutput=1
\usepackage{graphicx} 
\usepackage{jcappub}
\usepackage{xcolor}
\usepackage{graphicx}
\usepackage{float}
\usepackage{wrapfig}
\usepackage{bm}
\usepackage{ulem}

\usepackage{amsmath,amssymb}
\usepackage{subcaption}
\usepackage{verbatim}
\usepackage{makeidx}
\usepackage{braket}
\usepackage{multirow}
\usepackage[utf8]{inputenc}

\def\apj{\rm{ApJ}}

\def\mnras{\rm{MNRAS}}             

\newcommand*\diff{\mathop{}\!\mathrm{d}}
\newcommand*\Diff[1]{\mathop{}\!\mathrm{d^#1}}
\renewcommand\vec{\mathbf}

\graphicspath{{./Figures/}}

\title{Exploring compensated isocurvature perturbations with CMB spectral distortion anisotropies}

\author[a]{Taku Haga}
\emailAdd{haga@th.phys.titech.ac.jp}
\affiliation[a]{Department of Physics, Tokyo Institute of Technology,\\
Tokyo 152-8551, Japan}

\author[b,c]{Keisuke Inomata}
\affiliation[b]{ICRR, The University of Tokyo, Kashiwa, Chiba 277-8582, Japan}
\affiliation[c]{Kavli IPMU (WPI), UTIAS, The University of Tokyo, Kashiwa, Chiba 277-8583, Japan}
\emailAdd{inomata@icrr.u-tokyo.ac.jp}

\author[a,d]{Atsuhisa Ota}

\affiliation[d]{Institute for Theoretical Physics and Center for Extreme Matter and Emergent Phenomena,
Utrecht University,\\ Leuvenlaan 4, 3584 CE Utrecht, The Netherlands}
\emailAdd{a.ota@uu.nl}

\author[e,f]{and Andrea Ravenni}
\affiliation[e]{Dipartimento di Fisica e Astronomia “G. Galilei”, Universit{\`a} degli Studi di Padova, via~Marzolo~8, I-35131, Padova, Italy}
\affiliation[f]{INFN, Sezione di Padova, via~Marzolo~8, I-35131, Padova, Italy}
\emailAdd{ravenni@pd.infn.it}

\abstract{We develop a linear perturbation theory for the spectral $y$-distortions of the cosmic microwave background~(CMB).
The $y$-distortions generated during the recombination epoch are usually negligible because the energy transfer due to the Compton scattering is strongly suppressed at that time,  but they can be significant if there is a considerable amount of compensated isocurvature perturbation~(CIP), which is not tightly constrained from the present CMB observations.
The linear $y$-distortions explicitly depend on the baryon density fluctuations, therefore $y$ anisotropies can completely resolve the degeneracy between the baryon isocurvature perturbations and the cold dark matter ones.
This novel method is free from lensing contaminations that can affect the previous approach to the CIPs based on the nonlinear modulation of the CMB anisotropies. 
We compute the cross correlation functions of the $y$-distortions with the CMB temperature and the $E$ mode polarization anisotropies.
They are sensitive to the correlated CIPs parameterized by $f'\equiv\mathcal P_{\rm CIP\zeta}/\mathcal P_{\zeta \zeta}$ with $\mathcal P_{\zeta \zeta}$ and $\mathcal P_{\rm CIP\zeta}$ being the auto correlation of the adiabatic perturbations and the cross correlation between them and the CIPs.
We investigate how well the $y$ anisotropies will constrain $f'$ in future observations such as those provided by a PIXIE-like and a PRISM-like survey, LiteBIRD and a cosmic variance limited~(CVL) survey,
taking into account the degradation in constraining power due to the presence of Sunyaev Zel'dovich effect from galaxy clusters.
For example, our forecasts show that it is possible to achieve an upper limit of $f'< 2 \times 10^{5}$ at 68\% C.L. with LiteBIRD, and $f'<2\times 10^{4}$ with CVL observations.
}


\begin{document}
\begin{flushright}
IPMU18-0078
\end{flushright}
\maketitle
 
\section{Introduction}

The observation of the cosmic microwave background~(CMB) strongly suggests that the primordial density perturbations were adiabatic;
this implies that the number densities of photons~($\gamma$), baryons~($b$), cold dark matter~(CDM or $c$ for short), and neutrinos~($\nu$) fluctuate in the same way~\cite{Ade:2015xua}.
Departure from adiabatic perturbations,
such as additional isocurvature perturbations
on top of the adiabatic perturbations, change the CMB temperature angular powerspectrum drastically.
Therefore, we can set stringent constraints on isocurvature perturbations.
However, the current constraints have an exact degeneracy between baryon isocurvature perturbations and CDM ones~\cite{Gordon:2002gv}.
The difference between these two sectors appears in the linearized continuity equations for the baryons and the photons, because
only the baryon sector has additional terms proportional to the sound velocity squared $c_{s}^{2}$.
During the recombination period, we approximately have $c^{2}_{s}\approx \bar \rho_{\gamma}/\bar\rho_{b}
\approx (T_{\gamma}/m_{\rm p})\approx 10^{-9} $ with $T_{\gamma}$ and $m_{\rm p}$ being the temperature of photon at recombination and the proton mass, respectively~\cite{Grin:2011tf}. Then, the fluctuations of each sector follow the exact same equations before the sound horizon entry, and hence we cannot make a clear distinction between these two isocurvature modes.
For the temperature angular powerspectrum, the above Jeans scale corresponds to the multipole $\ell\sim 10^{6}$, where the anisotropies are exponentially suppressed due to the diffusion process called Silk damping.
Thus studying linear anisotropies, we can only constrain the total matter isocurvature and neutrino isocurvature perturbations.
The isocurvature perturbations of the single species can be combined in such a way that they leave no imprint on the linear CMB powerspectrum.
In this case they are called compensated isocurvature perturbations~(CIPs)~\cite{Gordon:2002gv}, which could possibly be related to the curvaton model~\cite{He:2015msa} or the spontaneous baryogenesis~\cite{DeSimone:2016ofp}.

Thus, the CIPs are less constrained compared to the observed adiabatic perturbations and the other isocurvature perturbations.
The current upper bounds on CIPs are given by looking at the secondary modulation of the baryon~(electron) distribution in the presence of CIPs at recombination~\cite{Grin:2011tf,Munoz:2015fdv,He:2015msa,Heinrich:2017psm,Valiviita:2017fbx}.
The amplitude of the dimensionless CIP powerspectrum $\Delta_{\rm rms}^{2}$ has been constrained with the \textit{Planck} data using different methodologies, such as looking for modifications of the CMB angular powerspectrum due to second order contributions, or through the analysis of the CMB trispectrum, whereas CIPs and the lensing effect are degenerate.
The most recent upper limit have been given, for example, in \cite{Munoz:2015fdv}, $\Delta_{\rm rms}^{2}<5.0\times 10^{-3}$ (68\% C.L.); in \cite{Valiviita:2017fbx} $\Delta_{\rm rms}^{2}<1.2\times 10^{-2}$ (95\% C.L.); and \cite{Grin:2013uya} $\Delta_{\rm rms}^{2}<1.1\times 10^{-2}$ (95\% C.L.).
As pointed out in Refs.~\cite{He:2015msa,Heinrich:2017psm}, further understanding is expected from the data analysis of a cosmic variance limited~(CVL) CMB-polarization survey, such as CMB-S4~\cite{Abazajian:2016yjj}.
There are also studies about the effects of CIPs on the baryon acoustic oscillation~\cite{Soumagnac:2016bjk,Soumagnac:2018atx}.
Another possibility is to see directly the spatial electron distribution by observing the redshift of 21 cm line of the neutral hydrogen hyperfine structure~\cite{Gordon:2009wx}.
This method is simple to discuss the hidden baryon fluctuations, but the actual detection of the signals would be challenging.
Most studies so far only provided upper limit to the existence of CIPs.
The author in Ref.~\cite{Valiviita:2017fbx} instead has recently reported a $2\sigma$ detection of the CIPs as the result of a full parameter estimation of Planck data, providing
a hint that a sizable amount of CIPs might exist.
Therefore, it will be interesting to pursue this trail using future cosmological data, and finding novel probes of CIPs signatures.

In this paper, we explore the possibility of observing the CIPs using CMB spectral distortion anisotropies.
The spectrum of the CMB has been measured with great precision by COBE/FIRAS, and it perfectly follows a (direction dependent) Planck distribution, within experimental accuracy.
However, it is actually non trivial that the deviation from the isotropic blackbody spectrum can be described solely by the local temperature parameter since the early universe is out of thermal equilibrium at redshifts $z\sim \mathcal O(10^{3})$ when it is dominated by the Compton scattering with energy transfer between photons and electons~\cite{1977NCimR...7..277D,1991MNRAS.248...52B}.
Once we have some energy transfer between photons and electrons during this period, the CMB spectrum is no more at equilibrium.
Such an energy transfer can be characterized by $T_{\rm e}/m_{\rm e}$ with $T_{\rm e}$ and $m_{\rm e}$ being the electron temperature and electron mass.
This correction is so tiny that it is usually ignored in the standard cosmological perturbation theory, instead it would be non negligible if a considerable amount of CIP is present.
Distribution functions for non equilibrium systems are highly non trivial, but it is known that in the regime of inefficient Compton scattering the photon distribution has a characteristic shape.
This is called $y$ spectral distortion, and its background component can be given as~\cite{Zeldovich:1969ff,Sunyaev:1970er}

\begin{align}
y = \int^{\eta_{0}}_{0}d\eta\,  n_{\rm e} \sigma_{\rm T}a \frac{T_{\rm e}-T_{\gamma}}{m_{\rm e}}\ ,
\end{align}
where $n_{\rm e}$ is the electron number density, $\sigma_{\rm T}$ is the Thomson scattering cross section, $a$ is the scale factor, and $\eta$ is the conformal time.
We consider the linear modulation of this parameter due to the electron density fluctuations in the presence of CIPs.
The vital point is that linear anisotropies of the $y$-distortions can resolve the degeneracy between CDM and baryon isocurvature perturbation, even though the signal is tiny because the energy transfer is suppressed as we mentioned above.
It is also important that we can easily distinguish the CIPs from the lensing effect in this way.
Here, we investigate the spectral distortion anisotropies by employing a systematic formulation proposed in Refs.~\cite{Stebbins:2007ve,Pitrou:2014ota,Ota:2016esq}.
The idea of using spectral distortions has been originally proposed in Ref.~\cite{Chluba:2013dna}, but it was shown that the quadratic spectral distortions from Silk damping are insensitive to the CIPs.
Instead in this article we consider the linear modulation of the Background spectral distortions.

This paper is organized as follows.
In section~\ref{Review-of-linear}, we provide a review of the linear perturbation theory for the temperature perturbations to contrast it with that for the spectral distortions.
Then, in section~\ref{SDA} we derive the linear $y$-distortion collision terms for the Compton scattering by using a set of frequency basis, obtain the Boltzmann equation, 
and estimate the linear $y$ anisotropies by using the Boltzmann code \texttt{CLASS}~\cite{Blas:2011rf}.
A Fisher matrix analysis about the possibility of constraining the CIPs with future surveys is given in section~\ref{Fisher}, and finally we draw our conclusions in section \ref{sec:Conclusions}.

\section{Brief 
review on linear temperature perturbations
}
\label{Review-of-linear}

In this section, we briefly review the linear perturbation theory for the CMB temperature anisotropies in terms of the frequency structure to emphasize what is new in this paper.

\subsection{Thomson scattering dominant universe}

When one fits the observed CMB photon energy spectrum with a Planck distribution for each direction $\mathbf n$, she finds the temperature field $T(\eta, \mathbf x,\mathbf n)$, where $(\eta,\mathbf x)$ is the space-time comoving coordinate.
The dimensionless temperature fluctuations are defined using the temperature of a time independent reference black-body $T_{\rm rf}$ as
\begin{align}
\Theta(\eta,\mathbf x,\mathbf n) = \frac{T(\eta, \mathbf x,\mathbf n)-T_{\rm rf}}{T_{\rm rf}}\ .
\end{align}
We usually solve the following Boltzmann equation 
alongside those for the other matter contents, and the linearized Einstein equation:
\begin{align}
\dot \Theta +\mathbf n\cdot \nabla\Theta -\dot \phi+\mathbf n\cdot \nabla \psi= n_{\rm e}\sigma_{\rm T}a\left(\Theta_{0}-\Theta + \mathbf n\cdot \mathbf v - \frac{1}{2}P_{2}\Theta_{2}\right),\label{Bolt:lin}
\end{align}
where the overdot is the partial derivative in terms of $\eta$, $\nabla$ is the spatial gradient, $\mathbf v$ is the velocity of the baryon fluid.
We ignored the polarizations for simplicity.
We work in the conformal Newtonian gauge
\begin{align}
g_{00}&=-a^2(1+2\psi),\\
g_{0i}&=g_{i0}=0,\\
g_{ij}&=a^2(1-2\phi)\delta_{ij}.
\end{align}
We also introduced the multipole component of a field $X$ as
\begin{align}
X_{\ell}=\frac{1}{(-i)^{\ell}}\int \frac{d\lambda}{2}P_{\ell}(\lambda)X (\lambda),
\end{align}
where $\lambda$ is the cosine between $\mathbf v$ and $\mathbf n$, and the $P_{\ell}$ are the Legendre polynomials.
We derived Eq.~(\ref{Bolt:lin}) linearizing the photon Boltzmann equation
\begin{align}
\frac{df}{d\eta}=\mathcal C_{\rm T}[f]\, ,\label{Bolt:def}
\end{align}
where $\mathcal C_{\rm T}$ is the collision term for the Thomson scattering.
Eq.~(\ref{Bolt:def}) has an explicit frequency dependence which disappeared in Eq.~(\ref{Bolt:lin}).
In other words, for each $(\eta, \mathbf x, \mathbf n)$, Eq.~(\ref{Bolt:def}) represents a continuously infinite number of equations.
Eq.~(\ref{Bolt:def}) falls into the  single equation~(\ref{Bolt:lin})  because we implicitly imposed the following local equilibrium ansatz at the beginning:
\begin{align}
f(\eta,\mathbf x, p\mathbf n)=\frac{1}{e^{\frac{p}{T(\eta,\mathbf x,\mathbf n)}}-1}\,,\label{anz:loc}
\end{align}
where $p$ is the comoving frequency~(momentum).
Throughout this paper, we use the comoving coordinate to ignore the time evolution of the background quantities. 
We express this ansatz as ``local equilibrium'' since the distribution function is a Planckian when we fix the space-time coordinate and the direction of the line of sight.
Local equilibrium ansatz 
 is a bold assumption for
  a general solution to the Boltzmann equation, but we can justify this since we believe that at sufficiently early times the universe was in local thermal equilibrium.
It is easy to check that this ansatz works for the linearized Boltzmann equation as follows.
The time derivative on the left hand side~(LHS) of Eq.~(\ref{Bolt:def}) is
\begin{align}
\frac{df}{d\eta}=\frac{\partial f}{\partial \eta }+\frac{dx^{i}}{d\eta}\frac{\partial f}{\partial x^{i}}+\frac{dn^{i}}{d\eta }\frac{\partial f}{\partial n^{i}}+\frac{dp}{d\eta}\frac{\partial f}{\partial p}\,.
\end{align}
On the other hand, Eq.~(\ref{anz:loc}), up to linear order in $\Theta$, implies
\begin{align}
f=f^{(0)}+\left(-p\frac{\partial f^{(0)}}{\partial p}\right)\Theta +\cdots,\label{anz:loc:lin}
\end{align}
where we defined the distribution function of the reference black-body as 
\begin{align}
f^{(0)}(p)=\frac{1}{\exp(p/T_{\rm rf})-1}\,.
\end{align}
Then we easily find
\begin{align}
\frac{df}{d\eta}=\left(
\frac{d\Theta}{d\eta}-\frac{d \ln p}{d\eta}
\right)\left(-p\frac{\partial f^{(0)}}{\partial p}\right)+\cdots,
\label{Lio:lin}
\end{align}
where the dots indicate the higher order contributions.
The logarithmic derivative of the frequency in Newtonian gauge is given as~\cite{Ma:1995ey}
\begin{align}
\frac{d\ln p}{d\eta}=\dot \phi-\mathbf n\cdot \nabla \psi.
\end{align}
Thus $d\ln p/d\eta$ is $p$-independent for massless particles and therefore all the frequency dependence can be factorized in the form of Eq.~(\ref{Lio:lin}).
On the other hand, the linear order Thomson scattering collision term is given as~\cite{Dodelson:1282338}
\begin{align}
C_{\rm T}[f]=n_{\rm e}\sigma_{\rm T}a\left(f_{0}-f+\mathbf n\cdot \mathbf v\left(-p\frac{\partial f^{(0)}}{\partial p}\right)-\frac{1}{2}P_{2}f_{2}\right).\label{Col:Tom:lin}
\end{align}
Similarly to the Liouville terms, Eqs.~(\ref{anz:loc:lin}) and (\ref{Col:Tom:lin}) yield
\begin{align}
C_{\rm T}[f]=n_{\rm e}\sigma_{\rm T}a\left(\Theta_{0}-\Theta+\mathbf n\cdot \mathbf v-\frac{1}{2}P_{2}\Theta_{2}\right)\left(-p\frac{\partial f^{(0)}}{\partial p}\right).\label{Col:Tom:lin:fct}
\end{align}
Combining Eqs.~(\ref{Lio:lin}) with (\ref{Col:Tom:lin:fct}), one finds Eq.~(\ref{Bolt:lin}), or equivalently we obtain the following equation after integrating over the frequency:
\begin{align}
\dot F_{\gamma} +\mathbf n\cdot \nabla F_{\gamma} 
-4\dot \phi+4 \mathbf n\cdot \nabla \psi
= n_{\rm e}\sigma_{\rm T}a\left(F_{\gamma 0}-F_{\gamma} + 4\mathbf n\cdot \mathbf v - \frac{1}{2}P_{2}F_{\gamma 2}\right),\label{Bolt:dens:lin}
\end{align}
where we defined the photon density fluctuation as
\begin{align}
F_{\gamma} \equiv \frac{\int p^{2}dp p\left(f-f^{(0)}\right)}{\int p^{2}dp pf^{(0)}}\,.
\end{align}
$F_{\gamma}=4\Theta$ is established at linear order, and the second order version of Eq.~(\ref{Bolt:dens:lin}) is also obtained straightforwardly~\cite{Hu:1993tc,Dodelson:1993xz,Bartolo:2006cu}.
However, the frequency cancellation for Eqs.~(\ref{Lio:lin}) and (\ref{Col:Tom:lin:fct}) is not possible when we have next-to-leading 
order corrections.
This implies limitations of the local equilibrium ansatz~(\ref{anz:loc}), that is, the CMB is no more blackbody when we go to the next-to-leading order in the energy-exchange expansion of the Compton scattering process.

\subsection{Beyond the Thomson scattering limit}

Let us go beyond the Thomson scattering limit, that is, we consider the collision processes of the Compton scattering, which, contrary to the Thompson scattering limit, allow energy transfer between photons and electrons.
Though the Thomson collision effect does not have the homogeneous part at the zeroth order, this is not the case if we include the Compton scattering correction as explicitly shown below~\cite{Ko:1957aa}:
\begin{align}
&(n_{\rm e}\sigma_{\rm T}a)^{-1}\mathcal C^{(0)}_{\rm K}[f]=\notag \\
&\frac{ T_{\rm e}}{m_{\rm e}}\left( p^2\frac{\partial^2 f^{(0)}}{\partial p^2}+4 p\frac{\partial f^{(0)}}{\partial p}\right)
+\frac{p(1+z)}{m_{\rm e}}\left(2 f^{(0)}p\frac{\partial f^{(0)}}{\partial p}+p\frac{\partial f^{(0)}}{\partial p}+4f^{(0)}{}^2+4f^{(0)}\right)\label{CK:0}.
\end{align}
Apparently, the frequency dependence cannot be summarized as we do at linear order.

The linear collision term for the Compton scattering is more complicated
since we need to expand the collision terms up to the cubic order in the electron velocity. 
The statistical average of the electron momentum cube can be divided into a thermal part that goes like $T_{\rm e}/m_{\rm e}\times v$ and a bulk velocity part such as $v^{3}$, which we drop for simplicity. 
We can drop the second order terms such as $v^{2}$ due to the Gaussian initial condition we assume.
This point will be discussed later.
The expression was first derived in Ref.~\cite{Chluba:2012gq} and rederived in Ref.~\cite{Ota:2016esq}, and has the form
\begin{align}
&(n_{\rm e}\sigma_{\rm T}a)^{-1}\mathcal C^{(1)}_{\rm K}[f]\notag \\
&=\frac{p(1+z)}{m_{\rm e}} \left[2p \frac{\partial f^{(0)}}{\partial p} f^{(1)}(\mathbf p)+4 f^{(0)} f^{(1)}(\mathbf p)+2 f^{(1)}(\mathbf p)
\right.\notag \\
&
\left.+4 f^{(0)} f^{(1)}_0  +p \frac{\partial f^{(1)}_0}{\partial p}+2 f^{(1)}_0+2 f^{(0)} p\frac{\partial f^{(1)}_0}{\partial p}\right.
\notag\\
&\left.
+(\mathbf{\hat v}\cdot \mathbf n)\left(\frac{24}{5} i f^{(0)} f^{(1)}_1+\frac{12}{5} i f^{(1)}_1+\frac{12i}{5} f^{(0)} p\frac{\partial f^{(1)}_1}{\partial p}+\frac{6 i}{5} p \frac{\partial f^{(1)}_1}{\partial p}\right)\right.\notag \\
&
\left.
+(\mathbf v\cdot \mathbf n) \left(-8 f^{(0)}{}^2-8 f^{(0)}-\frac{7}{5} p^2\frac{\partial^2 f^{(0)}}{\partial p^2}-\frac{14}{5}  f^{(0)} p^2\frac{\partial^2 f^{(0)}}{\partial p^2}-\frac{31}{5} p \frac{\partial f^{(0)}}{\partial p}-\frac{62}{5}  f^{(0)}p \frac{\partial f^{(0)}}{\partial p}\right)\right.\notag \\
&\left.
+P_2 \left(-2 f^{(0)} f^{(1)}_2-f^{(1)}_2- f^{(0)} p\frac{\partial f^{(1)}_2}{\partial p}-\frac{1}{2}p\frac{\partial f^{(1)}_2}{\partial p}\right)\right.\notag \\
&\left.  
+P_3 \left(-\frac{3i}{5}  f^{(1)}_3-\frac{6i}{5} f^{(0)} f^{(1)}_3-\frac{3 i}{5} f^{(0)} p \frac{\partial f^{(1)}_3}{\partial p}-\frac{3 i}{10}  p\frac{\partial f^{(1)}_3}{\partial p}\right)\right]
\notag \\
&
+\frac{T_{\rm e}}{m_{\rm e}} \left[(\mathbf v\cdot \mathbf n) \left(-\frac{7}{5} p^3 \frac{\partial^3 f^{(0)}}{\partial p^3}-\frac{47}{5} p^2 \frac{\partial^2 f^{(0)}}{\partial p^2}-\frac{15}{2} p \frac{\partial f^{(0)}}{\partial p}\right)+\right.\notag \\
&\left.
(\mathbf{\hat v}\cdot \mathbf n)  \left(\frac{6i}{5}  p^2 \frac{\partial^2 f^{(1)}_1}{\partial p^2}+\frac{24i}{5}  p \frac{\partial f^{(1)}_1}{\partial p}+\frac{6i}{5}  f^{(1)}_1\right)+P_2 \left(-\frac{1}{2} p^2 \frac{\partial^2 f^{(1)}_2}{\partial p^2}-2 p \frac{\partial f^{(1)}_2}{\partial p}+3 f^{(1)}_2\right)\right.\notag \\
&\left.
+P_3 \left(-\frac{3i}{10} p^2 \frac{\partial^2 f^{(1)}_3}{\partial p^2}-\frac{6i}{5}  p \frac{\partial f^{(1)}_3}{\partial p}+\frac{6i}{5}  f^{(1)}_3\right)+p^2 \frac{\partial^2 f^{(1)}_0}{\partial p^2}+4 p \frac{\partial f^{(1)}_0}{\partial p}\right]\label{CK:1},
\end{align}
where $f^{(1)}$ implies the linear fluctuations of the photon distribution function, 
which cannot be expressed solely in terms of temperature perturbation as we will see below.
Obviously, frequency dependence for the linear Compton collision term cannot be treated in a simple manner.
This clearly shows that our local equilibrium ansatz~(\ref{anz:loc:lin})
is no more applicable, and we only have the effective temperature perturbation
\begin{align}
\Theta(\eta,\mathbf x,\mathbf n)\to \Theta(\eta,\mathbf x,p\mathbf n)\,.
\end{align}
In principle we have to solve an infinite number of equations for each $(\eta,\mathbf x,\mathbf n)$.
This would be time consuming and requires tough numerical simulations.
In the next section, we solve this problem by employing a moment expansion, which was introduced in Refs.~\cite{Stebbins:2007ve,Pitrou:2014ota,Ota:2016esq}.

\section{Linear spectral distortions anisotropy}
\label{SDA}

The leading order Boltzmann equation falls into a single equation for the local temperature field.
It would be unnatural if the number of equations drastically changes from one to $\infty$ only because we included the next-to-leading order corrections.
The same problem also arises in the standard electrostatics.
Suppose we have a complicated static configuration of electric charges.
As long as we observe the potential field from afar, it can be approximated by a Coulomb potential, and the complicated structure appears in the multipole expansion order by order.
A similar method could be applied to the photon Boltzmann equation.
Suppose we consider the isotropic blackbody as the monopole, the temperature perturbations would be the dipole.
What corresponds to the quadrupole and the octupole?
What is the basis for the present moment expansion?

\subsection{Momentum expansion}

Unfortunately, we do not have a complete system of orthogonal polynomials such as the Legendre polynomial for the electrostatics.
Instead, Ref.~\cite{Ota:2016esq} found out that the following three basis functions would be minimum to handle the frequency dependence of the $T_{\rm e}/m_{\rm e} \to 0$ limit Boltzmann equation up to cubic order in the primordial fluctuations:
 \begin{align}
 \mathcal G \equiv  &   \left(-p \frac{\partial}{\partial p} \right) f^{(0)}, \label{def:G}
 \\
  \mathcal Y \equiv  &  \left(-p \frac{\partial}{\partial p} \right)^2 f^{(0)} - 3 \mathcal G, \label{def:Y} \\
 \mathcal K \equiv & \left(-p \frac{\partial}{\partial p} \right)^3 f^{(0)} - 3 \mathcal Y - 9 \mathcal G.\label{def:K}
     \end{align}
The following relations for the basis functions will be useful throughout this paper:
 \begin{align}
 \begin{split}
    p^2  \frac{\partial^2}{\partial p^2} f^{(0)} 
   & =  \mathcal Y + 4 \mathcal G ,\\   
   p^3  \frac{\partial^3}{\partial p^3} f^{(0)} 
   & = -\mathcal K - 6 \mathcal Y -20 \mathcal G .  
 \end{split}
 \label{UE:1}
      \end{align}
On the other hand, $T_{\rm e}/m_{\rm e}$ corrections make the problem more complicated in principle, but we newly found that adding the following frequency function is a minimal revision for the present linear case:
\begin{align}
  \mathcal U 
  &\equiv \frac{p^2}{T_{\rm rf}^2}  \left(-p \frac{\partial}{\partial p} \right) f^{(0)}.\label{def:U}
   \end{align}
Using the above frequency basis, let us consider the following ansatz:
\begin{align}
f =f^{(0)}+\Theta \mathcal G + y \mathcal Y + \kappa \mathcal K + u\mathcal U.\label{new:ans}
\end{align}
We will justify this ansatz in the later calculations.
As we already stated, the temperature perturbation is comparable to the order of the primordial fluctuation
that is denoted by $\delta $: $\Theta=\mathcal O(\delta )$.
On the other hand, the other coefficients would be the first order in both the energy transfer $\varepsilon = \mathcal O(T_{\rm e}/m_{\rm e})$ and 
$\delta $: $\{y,\kappa,u\}=\mathcal O(\delta  \varepsilon )$.

\subsubsection*{The homogeneous component}

Before going to the main discussion, we comment on the homogeneous part of the Compton scattering.
It is well known that Eq.~(\ref{CK:0}) can be recast into a simple form:
\begin{align}
(n_{\rm e}\sigma_{\rm T}a)^{-1}\mathcal C^{(0)}_{\rm K}[f]=\frac{ T_{\rm e}-T_{\gamma}}{m_{\rm e}}\mathcal Y,\label{Komp:homo}
\end{align}
where we have defined the physical temperature of photons as $T_{\gamma}\equiv T_{\rm rf}(1+z)$.
Note that we used Eq.~(\ref{UE:1}) with the following relations:
 \begin{align}
 \begin{split}
      f^{(0)} (1+ f^{(0)} )
   & = \frac{T_{\rm rf}}{p} \mathcal G, \\
    \left(-p \frac{\partial}{\partial p} \right) f^{(0)} (1+ f^{(0)} )
   & =\frac{T_{\rm rf}}{p} ( \mathcal Y + 4\mathcal G), \\
  \left(-p \frac{\partial}{\partial p} \right)^2 f^{(0)} (1+ f^{(0)} )
   & = \frac{T_{\rm rf}}{p} ( \mathcal K + 5 \mathcal Y + 16\mathcal G).
   \end{split}\label{UE:2}
   \end{align}
Eq.~(\ref{Komp:homo}) implies that $y$ has $O(\varepsilon)$ contribution
\begin{align}
y= \int d\eta n_{\rm e}\sigma_{\rm T}a \frac{ T_{\rm e}-T_{\gamma}}{m_{\rm e}}+\cdots.
\end{align}
However, this term is not what we want.
We consider the CIPs, which produce  the linear anisotropies of the spectral distortions, and hence we ignore this contribution.

\subsubsection*{The linear component}

In addition to Eq.~(\ref{CK:1}), we include the electron number density fluctuation on the top of homogeneous $y$-distortion in the previous subsection.
This is done by replacing the electron temperature and the number density as
\begin{align}
T_{\rm e}\to T_{\rm e}(1+\Theta_{\rm e}),~n_{\rm e}\to n_{\rm e}(1+\delta_{n_{\rm e}}),\label{mod:num:dens}
\end{align}
where $\delta_{n_{\rm e}}$ is the ionized electron number density perturbation and $\Theta_{\rm e}$ is the electron temperature perturbation.
We also solve the perturbed recombination for the evolution of $\delta_{n_{\rm e}}$ and $\Theta_{\rm e}$ following the equations in Ref.~\cite{Senatore:2008vi}.\footnote{
The perturbed recombination is discussed also in Refs.~\cite{Naoz:2005pd,Novosyadlyj:2006fw,Lewis:2007kz,Lewis:2007zh}.
}
Note that the equations for $\delta_{n_{\rm e}}$ and $\Theta_{\rm e}$ in Ref.~\cite{Senatore:2008vi} are valid only up to reionization and the precise calculation of $\delta_{n_{\rm e}}$ and $\Theta_{\rm e}$ during reionization is beyond the scope of this paper.\footnote{
We have confirmed that the reionization effect on $y$-distortions is subdominant at least in our setup.
}

Here, we comment on the initial conditions for the perturbed recombination.
We set $\delta_{n_{\rm e}}=\delta_b$ ($\delta_b$: baryon density perturbation) and $\Theta_{\rm e}=\Theta_{0}$ since we consider the photon baryon plasma to be in equilibrium state in the very early universe.
Note that $\Theta_{\rm e}=\delta_{b}/3$ is not necessarily established in the very early universe if we do not start with adiabatic perturbations.
For example, in the case of baryon isocurvature perturbations, $\delta_b$ depends on not only $\Theta_{\rm e}$ but also the chemical potential, and the spatial fluctuation of the chemical potential corresponds to the baryon isocurvature perturbations.
Therefore $\Theta_{\rm e}=\delta_{b}/3$ is not established in the case of the baryon isocurvature perturbations.

From Eq.~(\ref{CK:1}) and the modification (\ref{mod:num:dens}), we obtain the following collision terms for the Compton scattering:
\begin{align}
&(n_{\rm e}\sigma_{\rm T}a)^{-1}\mathcal C^{(1)}_{\rm C}[f]=\notag \\
&\frac{ T_{\rm e}(\Theta_{\rm e}+\delta_{n_{\rm e}}) }{m_{\rm e}}\left( p^2\frac{\partial^2 f^{(0)}}{\partial p^2}+4 p\frac{\partial f^{(0)}}{\partial p}\right)\notag \\
&
+\frac{p(1+z) \delta_{n_{\rm e}}}{m_{\rm e}}\left(2 f^{(0)}p\frac{\partial f^{(0)}}{\partial p}+p\frac{\partial f^{(0)}}{\partial p}+4f^{(0)}{}^2+4f^{(0)}\right)\notag \\
&+\frac{p(1+z)}{m_{\rm e}} \left[2p \frac{\partial f^{(0)}}{\partial p} f^{(1)}(\mathbf p)+4 f^{(0)} f^{(1)}(\mathbf p)+2 f^{(1)}(\mathbf p)
\right.\notag \\
&
\left.+4 f^{(0)} f^{(1)}_0  +p \frac{\partial f^{(1)}_0}{\partial p}+2 f^{(1)}_0+2 f^{(0)} p\frac{\partial f^{(1)}_0}{\partial p}\right.
\notag\\
&\left.
+(\mathbf{\hat v}\cdot \mathbf n)\left(\frac{24}{5} i f^{(0)} f^{(1)}_1+\frac{12}{5} i f^{(1)}_1+\frac{12i}{5} f^{(0)} p\frac{\partial f^{(1)}_1}{\partial p}+\frac{6 i}{5} p \frac{\partial f^{(1)}_1}{\partial p}\right)\right.\notag \\
&
\left.
+(\mathbf v\cdot \mathbf n) \left(-8 f^{(0)}{}^2-8 f^{(0)}-\frac{7}{5} p^2\frac{\partial^2 f^{(0)}}{\partial p^2}-\frac{14}{5}  f^{(0)} p^2\frac{\partial^2 f^{(0)}}{\partial p^2}-\frac{31}{5} p \frac{\partial f^{(0)}}{\partial p}-\frac{62}{5}  f^{(0)}p \frac{\partial f^{(0)}}{\partial p}\right)\right.\notag \\
&\left.
+P_2 \left(-2 f^{(0)} f^{(1)}_2-f^{(1)}_2- f^{(0)} p\frac{\partial f^{(1)}_2}{\partial p}-\frac{1}{2}p\frac{\partial f^{(1)}_2}{\partial p}\right)\right.\notag \\
&\left.  
+P_3 \left(-\frac{3i}{5}  f^{(1)}_3-\frac{6i}{5} f^{(0)} f^{(1)}_3-\frac{3 i}{5} f^{(0)} p \frac{\partial f^{(1)}_3}{\partial p}-\frac{3 i}{10}  p\frac{\partial f^{(1)}_3}{\partial p}\right)\right]
\notag \\
&
+\frac{T_{\rm e}}{m_{\rm e}} \left[(\mathbf v\cdot \mathbf n) \left(-\frac{7}{5} p^3 \frac{\partial^3 f^{(0)}}{\partial p^3}-\frac{47}{5} p^2 \frac{\partial^2 f^{(0)}}{\partial p^2}-\frac{15}{2} p \frac{\partial f^{(0)}}{\partial p}\right)+\right.\notag \\
&\left.
(\mathbf{\hat v}\cdot \mathbf n)  \left(\frac{6i}{5}  p^2 \frac{\partial^2 f^{(1)}_1}{\partial p^2}+\frac{24i}{5}  p \frac{\partial f^{(1)}_1}{\partial p}+\frac{6i}{5}  f^{(1)}_1\right)+P_2 \left(-\frac{1}{2} p^2 \frac{\partial^2 f^{(1)}_2}{\partial p^2}-2 p \frac{\partial f^{(1)}_2}{\partial p}+3 f^{(1)}_2\right)\right.\notag \\
&\left.
+P_3 \left(-\frac{3i}{10} p^2 \frac{\partial^2 f^{(1)}_3}{\partial p^2}-\frac{6i}{5}  p \frac{\partial f^{(1)}_3}{\partial p}+\frac{6i}{5}  f^{(1)}_3\right)+p^2 \frac{\partial^2 f^{(1)}_0}{\partial p^2}+4 p \frac{\partial f^{(1)}_0}{\partial p}\right].\label{Col:C:def}
\end{align}

The above expression is tedious but can be expanded in the four frequency basis functions Eqs.~(\ref{def:G}) to (\ref{def:U}).
Since Eq.~(\ref{Col:C:def}) is already ordered in terms of the Legendre polynomials, we easily obtain the multipole components of the Compton scattering for the Boltzmann hierarchy equation:

\begin{align}
\label{Col:Comp:0}
(n_{\rm e} \sigma_{\rm T} a)^{-1}\mathcal C_{\rm C0}^{(1)}[f]
 &= \left[\frac{T_{\rm e}}{m_{\rm e}} (\delta_{n_{\rm e}} +\Theta_{\rm e})- \frac{T_\gamma}{m_{\rm e}}(\delta_{n_{\rm e}} + \Theta_0) \right]\mathcal Y + \frac{T_{\rm e} - T_\gamma}{m_{\rm e}} \Theta_0 \mathcal K\, , \\
\label{Col:Comp:1}
(n_{\rm e} \sigma_{\rm T} a)^{-1}\mathcal C_{\rm C1}^{(1)}[f]
&=\left[
\frac{T_\gamma}{m_{\rm e}} 
\left( -\frac{1}{15} \Theta_1 - \frac{14}{45} iv \right)
+
 \frac{T_{\rm e}}{m_{\rm e}} 
 \left( -\frac{2}{5} \Theta_1 + \frac{7}{15} iv \right) 
 \right] \mathcal K 
 \nonumber\\ 
 &
+ \left[
\frac{T_\gamma}{m_{\rm e}} 
\left( \frac{2}{5} \Theta_1 + \frac{1}{5} iv \right)
+
 \frac{T_{\rm e}}{m_{\rm e}} 
\left( -\frac{1}{3} iv \right) 
 \right] \mathcal Y \nonumber\\ 
 &+ \left[
\frac{T_\gamma}{m_{\rm e}} 
\left( \frac{28}{15} \Theta_1 - \frac{28}{45} iv \right)
+
 \frac{T_{\rm e}}{m_{\rm e}} 
 \left( -\frac{2}{5} \Theta_1 - \frac{7}{10} iv \right) 
 \right] \mathcal G
 \nonumber\\ 
 &
+ \left[
\frac{T_\gamma}{m_{\rm e}} 
\left( \frac{7}{15} \Theta_1 - \frac{7}{45} iv \right)
\right] \mathcal U\, , \\
\label{Col:Comp:2}
(n_{\rm e} \sigma_{\rm T} a)^{-1}\mathcal C_{\rm C2}^{(1)}[f]
&=
\left(
-\frac{2}{5} \frac{T_\gamma}{m_{\rm e}} + \frac{1}{10} \frac{T_{\rm e}}{m_{\rm e}} 
\right) \Theta_2 \mathcal K
+
 \left(
-\frac{1}{10} \frac{T_\gamma}{m_{\rm e}} 
\right)  \Theta_2 \mathcal Y
\notag \\
&
+
\left(
\frac{6}{5} \frac{T_\gamma}{m_{\rm e}} - \frac{3}{5} \frac{T_{\rm e}}{m_{\rm e}} 
\right) \Theta_2 \mathcal G
+
 \left(
\frac{3}{10} \frac{T_\gamma}{m_{\rm e}} 
\right)  \Theta_2 \mathcal U\, ,
\\
\label{Col:Comp:3}
(n_{\rm e} \sigma_{\rm T} a)^{-1}\mathcal C_{\rm C3}^{(1)}[f]
&=
\left(
-\frac{32}{105} \frac{T_\gamma}{m_{\rm e}} - \frac{3}{70} \frac{T_{\rm e}}{m_{\rm e}} 
\right) \Theta_3 \mathcal K
+
 \left(
\frac{3}{70} \frac{T_\gamma}{m_{\rm e}} 
\right)  \Theta_3 \mathcal Y
\notag \\
&
+
\left(
\frac{146}{105} \frac{T_\gamma}{m_{\rm e}} + \frac{6}{35} \frac{T_{\rm e}}{m_{\rm e}} 
\right) \Theta_3 \mathcal G
+
 \left(
\frac{73}{210} \frac{T_\gamma}{m_{\rm e}} 
\right)  \Theta_3 \mathcal U\, ,
\end{align}
where $\mathcal C_{\rm C \ell}^{(1)}[f]$ is zero for $\ell>3$ and we used Eqs.~(\ref{UE:1}) and (\ref{UE:2}) with the following relations:
 \begin{align}
\left(- p \frac{\partial f^{(0)}}{\partial p} \right)^2
&= \frac{T_{\rm rf}}{6p} ( \mathcal K + 6 \mathcal Y +20 \mathcal G - \mathcal U),\label{UE:3}
\\
 p^2  \frac{\partial^2}{\partial p^2} \left(-p \frac{\partial}{\partial p} \right) f^{(0)}  
 &= \mathcal K +4 \mathcal Y +12 \mathcal G,
 \label{UE:4}
 \\
   p^2 \frac{\partial^2}{\partial p^2}  f^{(0)} (1+ f^{(0)} )
   &= \frac{T_{\rm rf}}{p} (\mathcal K + 6\mathcal Y + 20\mathcal G).
   \label{UE:5}
\end{align}

On the other hand, the LHS of the Boltzmann equation is simpler.
If one starts with the new ansatz~(\ref{new:ans}), Eq.~(\ref{Lio:lin}) is modified to 
\begin{align}
\frac{df}{d\eta}=\left(
\frac{d\Theta}{d\eta}-\frac{d \ln p}{d\eta}
\right)\mathcal G+ \frac{d y}{d\eta} \mathcal Y+ \frac{d \kappa}{d\eta} \mathcal K +\frac{d u}{d\eta} \mathcal U+ \cdots, 
\label{liu:new:ans}
\end{align}
where dots imply the next-to-leading order corrections.
Note that $d\ln p /d\eta=\mathcal O(\delta )$ so that the time derivatives of the frequency basis become at least linear order in the primordial perturbations.
As long as we consider Gaussian initial conditions, $\mathcal O(\delta^{2})$ can be dropped since these terms do not contribute to the cross correlations with the linear perturbations in the following discussions.
Though $\mathcal O(\delta^{3})$ has potential to be comparable to $\mathcal O(\delta \epsilon)$, we neglect these contributions for simplicity.
Combining Eqs.~(\ref{liu:new:ans}) with (\ref{Col:Comp:0}) to (\ref{Col:Comp:3}), we find the 4 coefficient equations for each $(\eta, \mathbf x)$.

\subsection{Boltzmann equation for the $y$-distortions}

We write the Boltzmann hierarchy equations in Fourier space to solve the linear $y$ evolution.
The Fourier integral of the real space linear $y$-distortions can be defined as
\begin{align}
y(\eta,\mathbf k,\mathbf n)\equiv \int d^{3}x e^{-i\mathbf k\cdot \mathbf x}y(\eta,\mathbf x,\mathbf n).
\end{align}
The Fourier space perturbations are linearly coming from the primordial perturbations.
The transfer functions $y^{(I)}_{\ell}(\eta,k)$ can be then given as
\begin{align}
y(\eta,\mathbf k,\mathbf n)=\sum_{\ell}(-i)^{\ell}(2\ell+1)P_{\ell}(\hat k\cdot \mathbf n)\sum_{I} y^{(I)}_{\ell}(\eta,k)\xi^{(I)}_{\mathbf k},
\label{eq:y_l_norm}
\end{align}
where $\xi^{(I)}$ is the scalar random variables, i.e., $\xi^{(I)}=(\zeta, S_{c\gamma}, S_{b\gamma}, \cdots)$: $\zeta$ is the adiabatic perturbation, and the isocurvature perturbations are defined as 
\begin{align}
S_{\alpha\gamma}\equiv \frac{1}{1+\omega_{\alpha}}\frac{\delta \rho_{\alpha}}{\rho_{\alpha}}
-\frac34\frac{\delta \rho_{\gamma}}{\rho_{\gamma}}.
\end{align}
Note that $\omega_{\gamma}=\omega_{\nu}=1/3$ and $\omega_{c}=\omega_{b}=0$.
We ignore the vector and the tensor perturbations for simplicity, and the following relation is satisfied for CIPs:
\begin{align}
S_{c\gamma}= - \frac{\Omega_b}{\Omega_c} S_{b\gamma}.\label{CIP:cond}
\end{align}

The Liouville terms for the linear $y$-distortion can be given in the same form with the temperature perturbations without metric perturbations.
Then Eqs.~(\ref{Col:Comp:0}) to (\ref{Col:Comp:3}) yield
 \begin{align}
\dot y_0 + k y_{1} &= n_{\rm e} \sigma_\text{T} a \left[ \frac{T_{\rm e}}{m_{\rm e}} (\delta_{n_{\rm e}} +\Theta_{\rm e})- \frac{T_\gamma}{m_{\rm e}}(\delta_{n_{\rm e}} + \Theta_0) \right],\label{hier:y:0}\\
\dot y_1 + \frac{2k}{3} y_2 - \frac{k}{3} y_0 &= -n_{\rm e} \sigma_\text{T} a y_1
+n_{\rm e} \sigma_\text{T} a \left[
\frac{T_\gamma}{m_{\rm e}} 
\left( \frac{2}{5} \Theta_1 + \frac{1}{5} iv \right)
+
 \frac{T_{\rm e}}{m_{\rm e}} 
\left( -\frac{1}{3} iv \right) 
 \right] ,\label{hier:y:1}\\
 \dot y_2 + \frac{3k}{5} y_3 - \frac{2k}{5} y_1 &= -\frac{9}{10} n_{\rm e} \sigma_\text{T} a y_2
+n_{\rm e} \sigma_\text{T} a
 \left(
-\frac{1}{10} \frac{T_\gamma}{m_{\rm e}} 
\right) \Theta_2,\\
\dot y_3 + \frac{4k}{7} y_4 - \frac{3k}{7} y_2 &= -n_{\rm e} \sigma_\text{T} a y_3
+n_{\rm e} \sigma_\text{T} a
 \left(
\frac{3}{70} \frac{T_\gamma}{m_{\rm e}} 
\right)  \Theta_3.
 \end{align}
$\ell>3$ equations do not have collision effect and we simply obtain~\cite{Dodelson:1282338}
\begin{align}
\dot y_\ell + \frac{\ell+1}{2\ell+1}k y_{\ell+1} - \frac{\ell}{2\ell+1} ky_{\ell-1} &= -n_{\rm e} \sigma_\text{T} a y_\ell \quad~(\ell>3).
\end{align}

\begin{figure}[htbp]
 \begin{minipage}{0.5\hsize}
  \begin{center}
   \includegraphics[width=73mm]{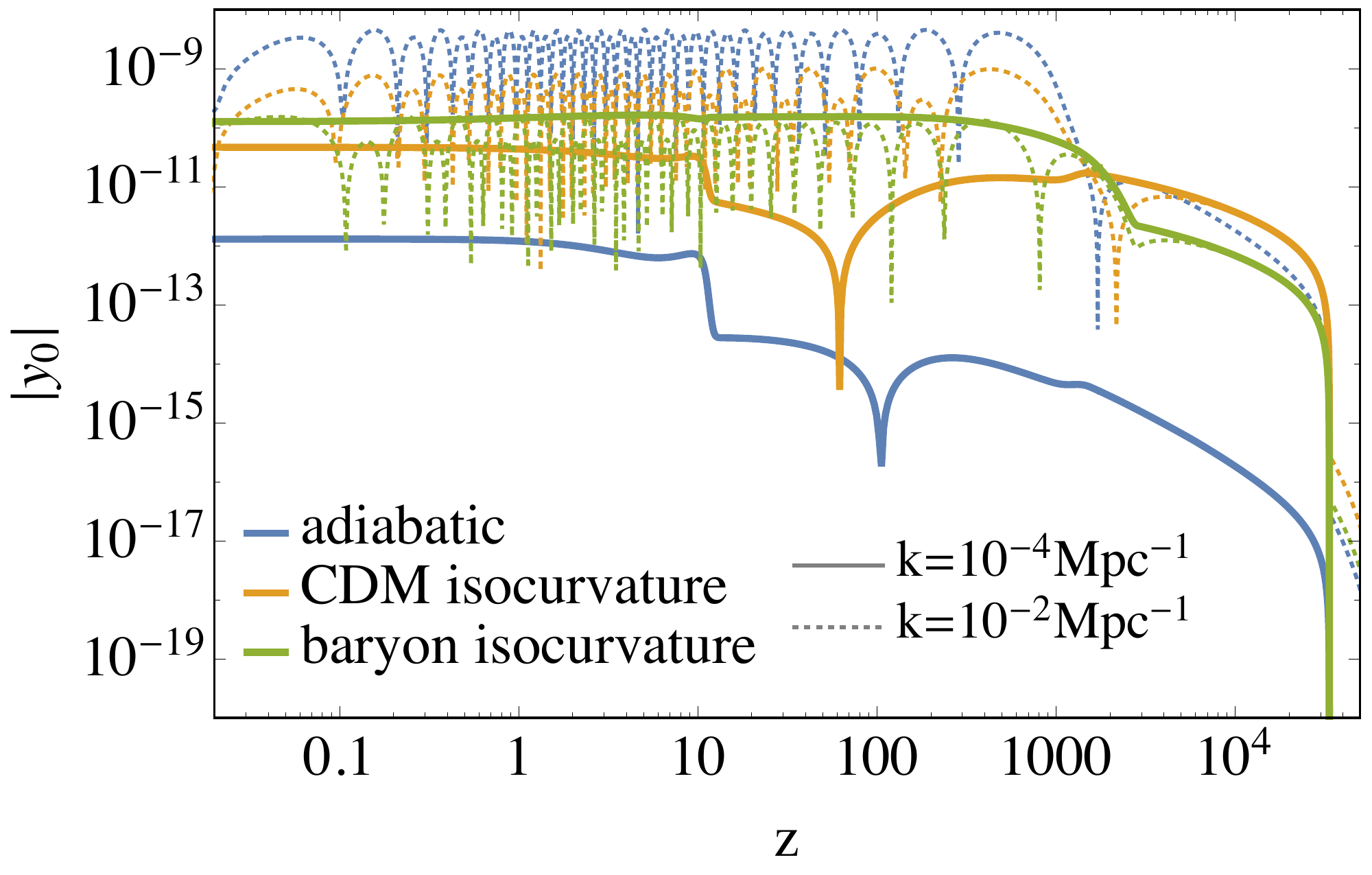}
  \end{center}
 \end{minipage}
 \begin{minipage}{0.5\hsize}
  \begin{center}
   \includegraphics[width=73mm]{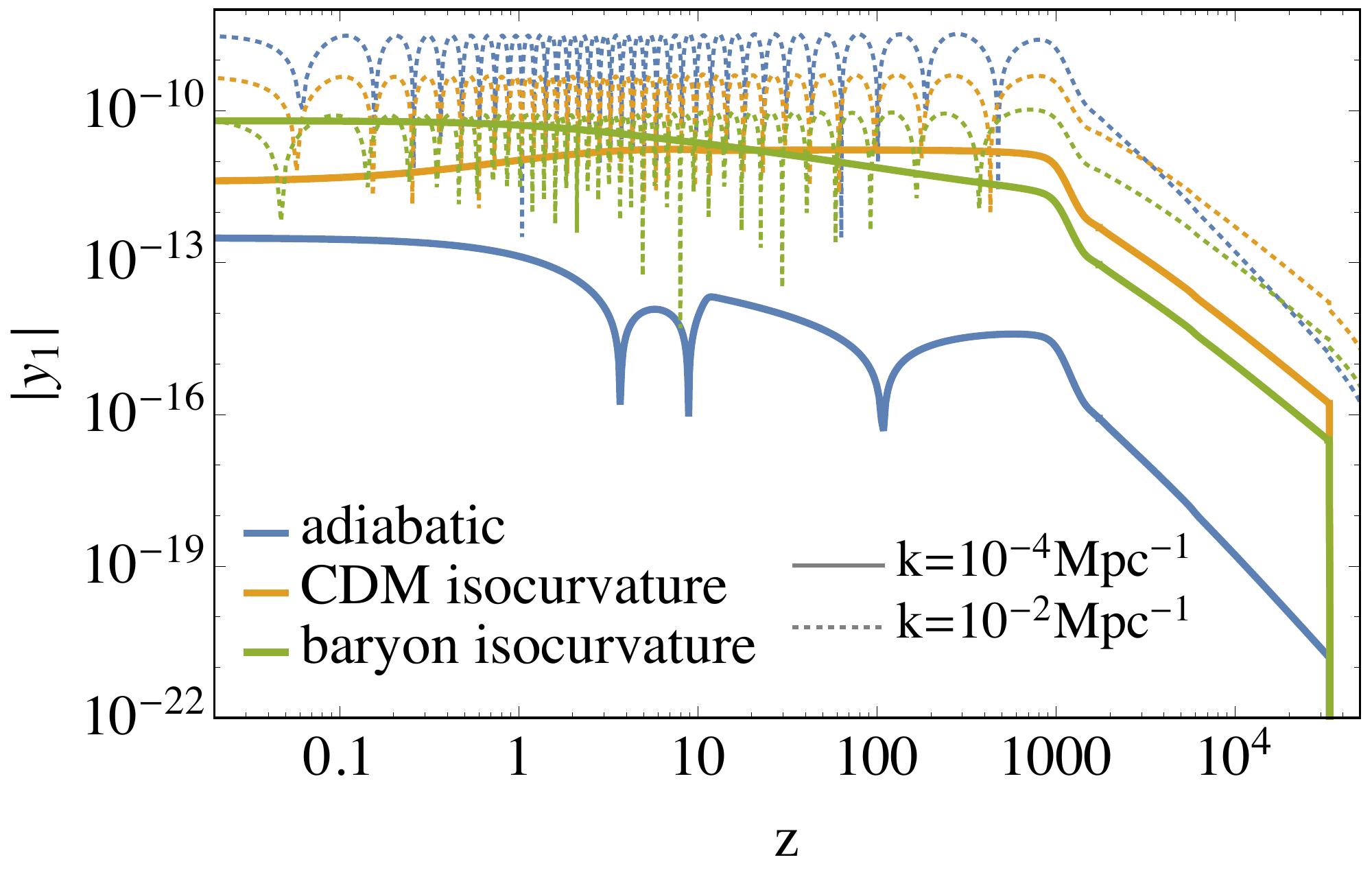}
  \end{center}
 \end{minipage}
 \caption{
 The redshift dependences of $|y_0|$ (left) and $|y_1|$ (right).
 Blue, green, yellow lines show the case of adiabatic perturbations, CDM isocurvature perturbations, and baryon isocurvature perturbations, respectively.
 In the solid (dotted) lines, we take the wave number as $k=10^{-4}$\,Mpc$^{-1}$ ($k=10^{-2}$\,Mpc$^{-1}$).
For example, blue solid line shows the case of the adiabatic perturbation with $k=10^{-4}$\,Mpc$^{-1}$.
 }
 \label{fig:y_evo}
\end{figure}

In the following we take synchronous gauge when we perform numerical calculations.
Figure~\ref{fig:y_evo} shows the evolutions of $|y_0|$ and $|y_1|$, which are calculated numerically using \texttt{CLASS}.
The rapid increases of values before $z=10^4$ is due to the specification of an initial time for integration.
We have checked that the final results are not sensitive to the choice of initial time as long as we start well before recombination.
From the figures, we can see that $|y_0|$ and $|y_1|$ grow only in the late epoch before recombination when the discrepancy between $T_{\rm e}$ and $T_{\gamma}$ becomes manifest.
This implies that the chemical potential 
$\mu$ distortions are not generated in this way.
After recombination, the sub-horizon modes oscillate 
 because the RHS of Eqs.~(\ref{hier:y:0}) and (\ref{hier:y:1}) is zero and so we have
\begin{align}
\ddot y_{0}-\frac{k^{2}}{3}y_{0}\approx 0.
\end{align}
Note that super-horizon evolution of the linear $y$-distortions does not violate 
causality.
The super-horizon $y$-distortions only appear along the super-horizon primordial fluctuations that already exist.

\subsection{Line-of-sight formalism for the spectral distortion anisotropies}

The remaining procedures to get the angular powerspectra are exactly the same as for the usual linear anisotropies.
The angular powerspectrum up to $\ell=1000$ requires us to solve 1000 Boltzmann hierarchy equations, which is time consuming.
Instead, we usually use the integral solution and solve a very limited number of 
hierarchy equations to obtain the source function for the integral solution~\cite{Seljak:1996is}.
The line-of-sight solution for the linear $y$-distortion is  
\begin{align}
 y(\eta_0,k, \lambda) = \int^{\eta_0}_0 d \eta \, g \left( y_0 - \frac{1}{2} P_2 y_2 + \mathcal S_{y} \right) e^{ik\lambda(\eta-\eta_0)},
\end{align}
where the multipole coefficients for the source function are 
\begin{align}
\mathcal S_{y0}&=\frac{T_{\rm e}}{m_{\rm e}} (\delta_{n_{\rm e}} +\Theta_{\rm e})- \frac{T_\gamma}{m_{\rm e}}(\delta_{n_{\rm e}} + \Theta_0)\, ,\\
\mathcal S_{y1}&= \frac{T_\gamma}{m_{\rm e}} 
\left( \frac{2}{5} \Theta_1 + \frac{1}{5} iv \right)
+
 \frac{T_{\rm e}}{m_{\rm e}} 
\left( -\frac{1}{3} iv \right) , \\
\mathcal S_{y2}&= 
-\frac{1}{10} \frac{T_\gamma}{m_{\rm e}} 
 \Theta_2\,, \\
\mathcal S_{y3}&= 
\frac{3}{70} \frac{T_\gamma}{m_{\rm e}} 
 \Theta_3\,.
\end{align}
Note that the baryon isocurvature perturbation dependences of $\delta_{n_{\rm e}}$ and $\Theta_{\rm e}$ are different from the CDM ones.
This is essential to distinguish the baryon isocurvature perturbations from the CDM ones, and therefore to observe the CIPs.
This is specific to $y$-distortions; $\Theta$, $\kappa$ and $u$ do not have such a source and therefore we only focus on the $y$-distortion linear anisotropies in this paper.
Note that the following form is 
more practical to use in \texttt{CLASS}:
\begin{align}
 y_{\ell}(\eta_0,k, \lambda)    = &\int^{\eta_0}_0 d \eta \, g \left[ y_0 + \mathcal S_{y0}
 + 3 \mathcal S_{y1} \left( \frac{\partial}{\partial (k \eta_{0})} \right) \right. \nonumber \\
 & + \left( 5 \mathcal S_{y2} + \frac{1}{2} y_2\right) \frac{1}{2} \left( 3 \left( \frac{\partial}{\partial (k \eta_{0})} \right)^2 +1 \right) \nonumber \\ 
  &
 \left. + 7 \mathcal S_{y3} \frac{1}{2} \left( 5  \left( \frac{\partial}{\partial (k \eta_{0})} \right)^3 + 3  \left( \frac{\partial}{\partial (k \eta_{0})} \right) \right)
  \right] j_{\ell}[k(\eta_{0}-\eta)]\, .
  \label{eq:y_eom_fourier}
\end{align}
The anisotropies of $X$ on the celestial sphere can be expanded in spherical harmonics as follows:
\begin{align}
 X(\eta_0,\mathbf x=0,\mathbf n) &= \sum^\infty_{\ell=0} \sum^\ell_{m=-\ell} a_{X,\ell m} Y_{\ell m}(\mathbf n), 
\end{align}
where $Y_{\ell m}$ are the spherical harmonics and $X=\Theta,~y$ and polarization $E$ mode.
The harmonic coefficients are related to the primordial random fields as
 \begin{align}
a_{X,\ell m} &= 4\pi (-i)^\ell  \int \frac{d^3 k}{(2\pi)^3} Y_{\ell m}^* ( \hat k) \sum_{I} X^{(I)}_\ell(\eta_0,k)\xi^{(I)}_{\mathbf k}.
 \label{eq:a_lm}
 \end{align}
The angular powerspectrum is also defined as in the usual linear anisotropies case:
 \begin{align}
\langle a_{X,\ell m} a^*_{Z,\ell' m'}\rangle = C^{XZ}_{\ell} \delta_{\ell \ell}\delta_{mm'}.
 \end{align}
Then, we find
\begin{align}
C^{XZ}_{\ell}= \sum_{II'}4\pi \int \frac{d k}{k} \mathcal P_{II'}(k) X^{(I)}_\ell(\eta_0,k) Z^{(I')}_\ell( \eta_0, k),
\label{eq:cl_yt_corr}
\end{align}
where we defined the powerspectra of the random variables as:
\begin{align}
\langle \zeta_{\mathbf k}\zeta_{\mathbf k'} \rangle &=(2\pi)^{3}\delta^{(3)}(\mathbf k +\mathbf k')\frac{2\pi^{2}}{k^{3}}\mathcal P_{\zeta\zeta}(k), \\
\langle \zeta_{\mathbf k}S_{\alpha\gamma, \mathbf k'} \rangle &=(2\pi)^{3}\delta^{(3)}(\mathbf k +\mathbf k')\frac{2\pi^{2}}{k^{3}}\mathcal P_{\zeta\alpha}(k),\\
\langle S_{\alpha\gamma, \mathbf k} S_{\beta\gamma, \mathbf k'} \rangle &=(2\pi)^{3}\delta^{(3)}(\mathbf k +\mathbf k')\frac{2\pi^{2}}{k^{3}}\mathcal P_{\alpha\beta}(k).
\end{align}
Note that the baryon isocurvature powerspectrum and the CDM isocurvature one  are not independent in the CIPs model: the relations 
\begin{align}
\mathcal P_{cc} = \left( \frac{\Omega_b}{\Omega_c} \right)^2 \mathcal P_{bb},~\mathcal P_{c\zeta} = -\frac{\Omega_b}{\Omega_c}\mathcal P_{b\zeta}\label{condCIP}
\end{align}
are always satisfied.

\begin{figure}
\centering
\includegraphics[width=0.7\textwidth]{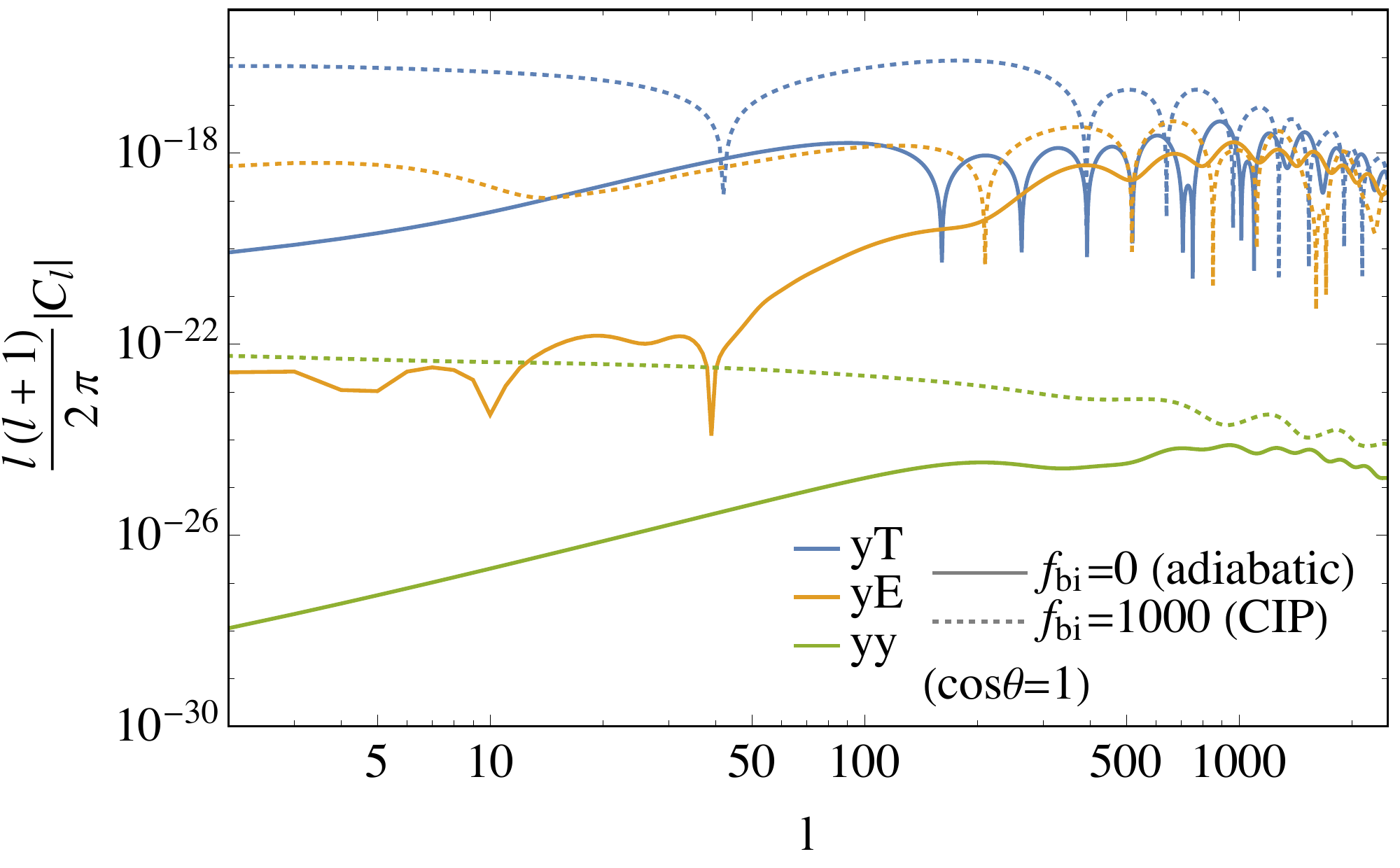}
\caption{$|C_{\ell}^{yT}|$, $|C_{\ell}^{yE}|$, and $|C_{\ell}^{yy}|$ in the case of adiabatic perturbations and CIP with $\cos \theta = 1$.
Blue, orange, and green lines show $|C_{\ell}^{yT}|$, $|C_{\ell}^{yE}|$, and $|C_{\ell}^{yy}|$
and solid and dotted lines show the case of $f_{\text{bi}}=0$ (corresponding to adiabatic) and $f_{\text{bi}}=1000$ (corresponding to CIP), respectively.
For example, the blue solid line shows $|C_{\ell}^{yT}|$ in the case of $f_{\text{bi}}= 0$.
 }
\label{fig:cl_summary}
\end{figure}

In this paper, we assume scale invariant CIPs for simplicity and normalize the powerspectrum by the adiabatic perturbations.
Then CIPs can be parameterized by only two parameters, $f_\text{bi}$ and $\cos \theta$, which are defined as
\begin{align}
\label{eq:cip_def_b}
f_\text{bi} &\equiv \sqrt{\frac{ \mathcal P_{bb}(k_0)}{\mathcal P_{\mathcal \zeta \zeta} (k_0)}}, \\
\cos \theta &\equiv \frac{  \mathcal P_{b \zeta }(k_0)} {\sqrt{\mathcal P_{bb} (k_0) \mathcal P_{\mathcal \zeta \zeta} (k_0)}},
\end{align}
where $k_0$($=0.05$\,Mpc$^{-1}$) is the pivot scale.
While $f_\text{bi}$ measure the amplitude of CIPs relative to the adiabatic perturbations, $\cos\theta$ parametrize the degree of correlation between the two kinds.
As we will discuss at length, our analysis is sensitive not only to $f_\text{bi}$, like other methods developed in the literature, but also to $\cos \theta$.
This would allow, in principle, the possibility of discerning correlated and uncorrelated CIPs.
Figure~\ref{fig:cl_summary} shows $C_{\ell}^{yT}$, $C_{\ell}^{yE}$, and $C_{\ell}^{yy}$ in both cases of adiabatic perturbations and CIPs.
From this figure, we can see that CIPs affect $C_\ell^{yT}$, $C_\ell^{yE}$, and $C_{\ell}^{yy}$ and there is a possibility to detect CIPs with the $y$-distortion anisotropies.

\label{subsec:cip}

\section{Fisher forecast}
\label{Fisher}

In this section we will investigate how well future CMB survey will constrain CIP 
 models, exploiting the $y$-distortions anisotropies.
To assess this we will produce Fisher forecasts for a PIXIE-like experiment \cite{Kogut:2011xw}, a PRISM-like experiment \cite{Andre:2013nfa}, for LiteBIRD \cite{Matsumura:2013aja} and for a Cosmic Variance Limited (CVL), very futuristic, experiment.
\\

Before discussing the Fisher matrix analysis, we explain the $f_{\rm bi}$ dependence of the angular powerspectra.
For the CIPs, $\Theta^{(c)}_{\ell}S_{c\gamma}+\Theta^{(b)}_{\ell}S_{b\gamma}\approx 0$ is satisfied so that angular power-spactra do not have auto correlation of the CIPs in Eq.~(\ref{eq:cl_yt_corr}) as follows:
\begin{align}
\label{eq:clyt_simple}
C^{yT}_{\ell} &\approx 4\pi \int \frac{d k}{k} \left( \mathcal P_{\zeta \zeta}(k) y^{(\zeta)}_\ell(\eta_0,k) \Theta^{(\zeta)}_\ell( \eta_0, k)  + \mathcal P_{c\zeta}(k) y^{({S_c})}_\ell(\eta_0,k) \Theta^{(\zeta)}_\ell( \eta_0, k) \right. \nonumber \\
&  \left. +  \mathcal P_{b\zeta}(k) y^{({S_b})}_\ell(\eta_0,k) \Theta^{(\zeta)}_\ell( \eta_0, k) \right),\\ 
\label{eq:clye_simple}
C^{yE}_{\ell} &\approx 4\pi \int \frac{d k}{k} \left( \mathcal P_{\zeta \zeta}(k) y^{(\zeta)}_\ell(\eta_0,k) E^{(\zeta)}_\ell( \eta_0, k)  + \mathcal P_{c\zeta}(k) y^{({S_c})}_\ell(\eta_0,k) E^{(\zeta)}_\ell( \eta_0, k) \right. \nonumber \\
&   \left. +  \mathcal P_{b\zeta}(k) y^{({S_b})}_\ell(\eta_0,k) E^{(\zeta)}_\ell( \eta_0, k) \right).
\end{align}
Using the angular powerspectra for the adiabatic perturbations $C_{\ell, \text{ad}}$ and those for the CIPs $C_{\ell, \text{CIP}}$ 
at $f_\text{bi} \cos \theta =1$, 
the above expressions are simply written as
\begin{align}
C^{yT}_{\ell} &\approx C_{\ell, \text{ad}}^{yT} + f_\text{bi} \cos\theta \,C_{\ell, \text{CIP}}^{yT}, \label{lin:yT:CIP}\\
C^{yE}_{\ell} &\approx C_{\ell, \text{ad}}^{yE} + f_\text{bi} \cos\theta \,C_{\ell, \text{CIP}}^{yE}.\label{lin:yE:CIP}
\end{align}

\begin{figure}
\centering
\includegraphics[width=0.7\textwidth]{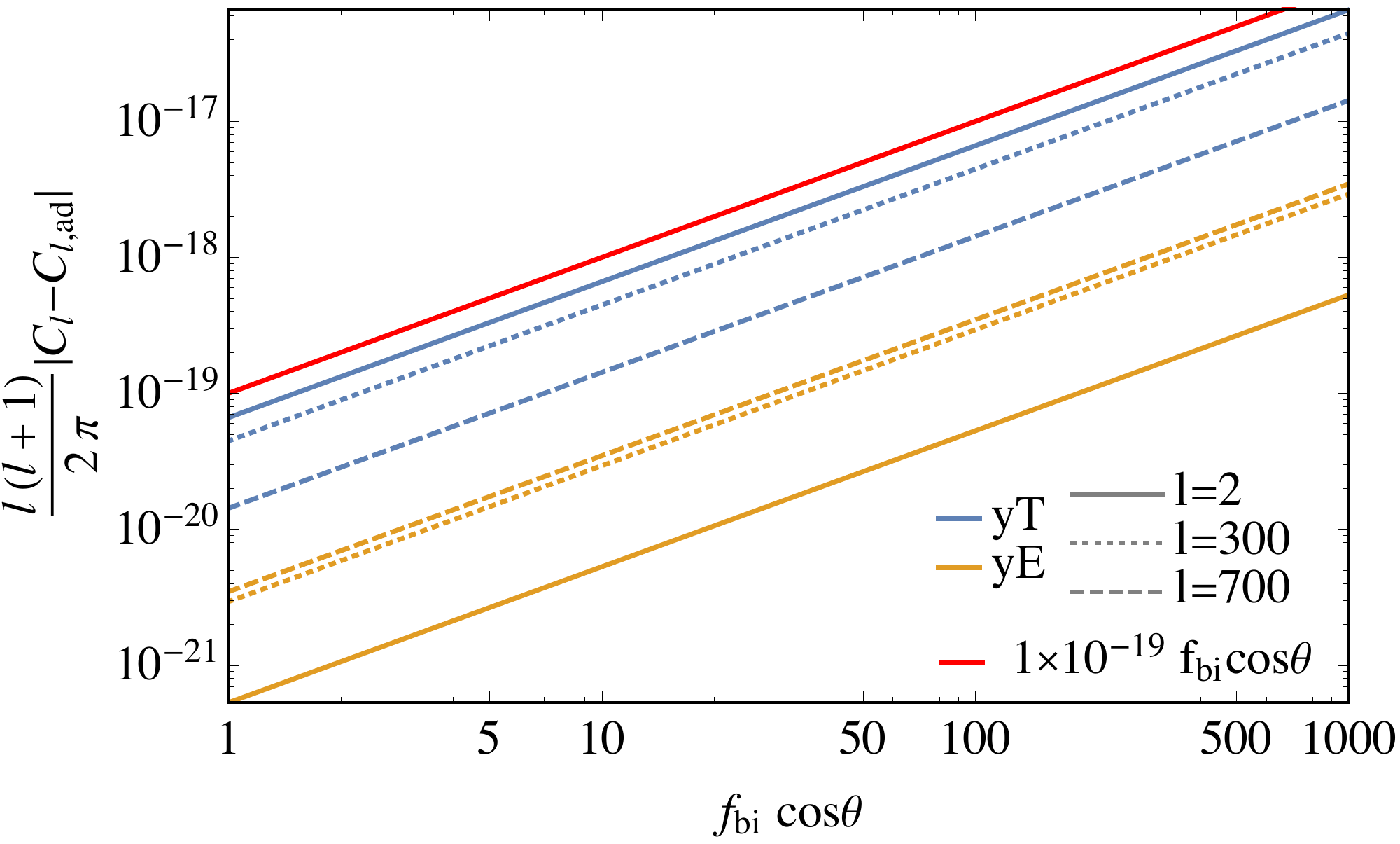}
\caption{$f_{\text{bi}}$ dependences of $|C_{\ell}^{yT}-C_{\ell, \text{ad}}^{yT}|$ and $|C_{\ell}^{yE}-C_{\ell, \text{ad}}^{yE}|$.
Blue and orange lines show $|C_{\ell}^{yT}-C_{\ell, \text{ad}}^{yT}|$ and $|C_{\ell}^{yE}-C_{\ell, \text{ad}}^{yE}|$
and solid, dotted and dashed lines show the case of $\ell=2$, $\ell=300$ and $\ell=700$, respectively.
For example, the blue solid line shows $|C_{\ell}^{yT}-C_{\ell, \text{ad}}^{yT}|$ in the case of $\ell=2$.
For comparison, we also plot $1\times 10^{-19} f_\text{bi} \cos \theta$ with a red line.
}
\label{fig:cl_fbi_dep}
\end{figure}

Thus, the angular powerspectra linearly depend on the correlated CIPs.
Figure~\ref{fig:cl_fbi_dep} shows the $f_\text{bi}\cos\theta$ dependence of $C_\ell-C_{\ell, \text{ad}}$ in $\ell=2$, $\ell=300$, and $\ell=700$.
From this figure, we can see that $C_\ell-C_{\ell, \text{ad}}$ is indeed proportional to $f_\text{bi} \cos \theta$.
Since the amplitudes of the $C_{\ell,\text{CIP}}^{yT}$ and $C_{\ell,\text{CIP}}^{yE}$ are proportional to the product of $f_\text{bi}$ and $\cos\theta$, our analysis will not be independently sensitive to each of these parameters, but we can test $f' \equiv f_\text{bi} \cos\theta$.
On the other hand, $C_{\ell,\text{CIP}}^{yy}$ is quadratic in $f_{\rm bi}$ but is dominated by astrophysical contamination discussed below.
\\

In the scope of our work, the Sunyaev-Zeldovic (SZ) effect constitutes a foreground.
Its powerspectrum adds an important contribution to the variance of the CIPs-generated $y$-T and $y$-E cross correlation.
We will show how masking resolved clusters \cite{Creque-Sarbinowski:2016wue, Ravenni:2017lgw} can greatly improve the constraining capabilities of all future surveys.
Moreover, the coupling of SZ effect with low redshifts sources of temperature (ISW) and polarization anisotropies could in principle bias the measurements of $y$-$T_{\rm CIP}$ $y$-$E_{\rm CIP}$, and therefore needs to be properly accounted for \cite{Ravenni:2017lgw}.
Luckily, the spectral shapes of the signals and of the spurious secondary sources are not much correlated.
Therefore marginalizing over them will not degrade the constraining power by much.
As long as we consider the Gaussian universe, other sources of $y$-distortions --- such as reionization, Silk damping --- contribute negligibly to the total $y$-$T$ and $y$-$E$ cross correlations \cite{Ravenni:2017lgw} therefore we will not consider them here.
In the case of primordial non-Gaussianity, it has been shown that foregrounds can degrade the constraints obtained with naive Fisher forecast by a factor $5 \sim 10$ \cite{Remazeilles:2018kqd}.
Given the similarity of our analysis with the search for primordial non-Gaussianity, we might expect a similar effect.
However, account properly for the galactic foregrounds would be beyond the (path-finding) scope of this paper.
Therefore we will neglect them completely in our analysis.
For this reason our results will be one order of magnitude too optimistic.

The Fisher matrix is defined as 
\begin{equation}
F_{ij}\equiv \Braket{ \frac{\partial L}{\partial{p_i}}\frac{\partial L}{ \partial{p_j}}} ,
\label{eq:Fdef}
\end{equation}
where $L$ is the logarithm of the likelihood and ${\bf p}$ are the free parameters of the theory.
In our application it is equivalent to \cite{Heavens:2009nx}
\begin{equation}
F_{ij}= f_\text{sky} \sum_\ell (\textbf{Cov}^{-1}_\ell)_{\alpha\beta} \frac{\partial (\textbf{Cov}_\ell)_{ \beta\gamma}}{\partial{p_i}}
(\textbf{Cov}_\ell^{-1})_{\gamma\delta} \frac{\partial (\textbf{Cov}_\ell)_{\delta\alpha}}{\partial{p_j}}\, ,
\label{eq:Foperative}
\end{equation}
where $f_\text{sky}$ is the fraction of the sky covered by the survey, $\textbf{Cov}_{\ell}$ is the covariance matrix of the observables and repeated matrix indices ($\alpha, \dots ,\delta$) are summed.
We will first discuss the results achieved considering only the cross correlation of $y$-distortions with temperature anisotropies (the same applies also to polarization).
The same approach will be then  
extended to the joint analysis of the cross correlation with both temperature and polarization anisotropies.
The covariance matrix of $(a_{\ell m}^T,a_{\ell m}^y)$ reads
\begin{equation}
\textbf{Cov}_{\ell}
=
\frac{1}{2\ell+1}\begin{pmatrix}
C_\ell^{TT}														& C_\ell^{y T} + C_\ell^{\text{SZ} T} \\
C_\ell^{y T} + C_\ell^{\text{SZ} T }   & C_\ell^{yy} + C_\ell^\text{SZSZ} + C_\ell^{yy,N} \\   
\end{pmatrix},
\label{eq:Tmarg}
\end{equation}
where we have explicitly separated the primordial component of the $y$-distortions from the signal coming from the SZ effect.
We have also considered its instrumental noise contribution  $C_\ell^{yy,N}$.

In principle all the six standard cosmological parameters and $f'$ need to be considered free parameters in the Fisher forecast.
However the uncertainties on the cosmological parameters are so small compared with the one of $f'$
that we consider them fixed to their true value when we calculate the temperature (and polarization) anisotropies.
This is a safe choice since $f'$ is completely orthogonal to the changes in $TT, TE$ and $EE$~\cite{Gordon:2002gv}. 
Uncertainties on $H_0$, $\sigma_8$, and $\Omega_m$ have a much bigger impact on the SZ powerspectrum.
Moreover, the halo mass bias $b$, which is the parameter that links the spectroscopic mass of the halo with its true mass (lensing mass), introduces another source of uncertainty in the SZ auto and cross-correlations.
As discussed in \cite{Aghanim:2015eva, Bolliet:2017lha, Salvati:2017rsn, Hurier:2017jgi}, the effects of varying these parameters are degenerate, and, on low $\ell$ ($\lesssim 1000$), can be parametrized as a shift in the overall amplitude of the SZ powerspectrum.
Following \cite{Creque-Sarbinowski:2016wue}, we will take into account the combined effect of these parameters by simply introducing an unknown amplitude parameter $\alpha$, in front of the SZ-spectra, and marginalizing over it.
We will calculate the SZ auto- and cross-correlation used here in the following section.
Using the covariance matrix (\ref{eq:Tmarg}) in Eq.~(\ref{eq:Foperative}) with Eq.~(\ref{lin:yT:CIP})
, we can obtain the Fisher matrix for the parameters ($f'$, $\alpha$)
\begin{equation}
F_{ij}= f_\text{sky} \sum_{\ell =2}^{\ell_\text{max}} (2\ell + 1)
\begin{pmatrix}
 \frac{ (C_{\ell,\text{CIP}}^{yT})^2}{C_\ell^{TT} (C_\ell^\text{SZSZ}+C_\ell^{yyN})} & \frac{ C_\ell^{\text{SZ}T} C_{\ell,\text{CIP}}^{yT}}{C_\ell^{TT} (C_\ell^\text{SZSZ}+C_\ell^{yyN})} \\
 \frac{ C_\ell^{\text{SZ}T} C_{\ell,\text{CIP}}^{yT}}{C_\ell^{TT} (C_\ell^\text{SZSZ}+C_\ell^{yyN})} & \frac{ (C_\ell^{\text{SZ}T})^2}{C_\ell^{TT} (C_\ell^\text{SZSZ}+C_\ell^{yyN})} \\
\end{pmatrix}.
\label{eq:FishT}
\end{equation}
To get this result we assumed that $(C_\ell^{\text{SZ}T})^2 \ll C_\ell^\text{SZSZ}C_\ell^{TT}$, and $C_\ell^{yy} \ll C_\ell^\text{SZSZ}$.
This treatment can be trivially generalized to produce joint analysis of $C_\ell^{yT}$ and $C_\ell^{yE}$, using the appropriate covariance matrix for $(a_{\ell m}^T,a_{\ell m}^E,a_{\ell m}^y)$:
\begin{equation}
 \textbf{Cov}_{\ell}
=
\frac{1}{(2\ell+1)}
\begin{pmatrix}
C_\ell^{TT}    & C_\ell^{TE}                                                                                 & C_\ell^{y T} + C_\ell^{\text{SZ} T}                                                       \\
 C_\ell^{TE}   & C_\ell^{EE}                                                                                 & C_\ell^{y E} + C_\ell^{\text{SZ} E} \\
 C_\ell^{y T} + C_\ell^{\text{SZ} T}& C_\ell^{y E} +C_\ell^{\text{SZ} E }  & C_\ell^{yy} + C_\ell^\text{SZSZ} + C_\ell^{yy,N}\\   
\end{pmatrix}.
\label{eq:FishTE}
\end{equation}

The temperature and polarization auto- and cross-correlation can be easily computed with \texttt{CLASS} \cite{Lesgourgues:2011re}, whereas the SZ powerspectrum and cross correlation with temperature and polarization anisotropies can be calculated using the halo model that we review below.

\subsection{SZ auto- and cross-correlations}

To calculate the SZ we use a halo model approach, following  \cite{2011MNRAS.418.2207T, Hill:2013baa, Komatsu:2002wc}.
We parametrize the density of dark matter halos in terms of the matter overdensity distribution $\delta$, using a halo bias parameter $b_H(z,M)$, which depends on redshift and mass of the halo.
The mass distribution of halos is given in terms of the halo mass function $\frac{\diff n}{\diff M}(z,M)$.
Since the SZ is sensitive to the electron distribution rather than to the matter distribution, this has to be convolved with the halo Compton $y$-parameter image $y_{3D}(z,M,x)$, where $x$ is the distance from the center of the halo; $y_{3D}$ is a function of the electron pressure profile of the halo.
We use the halo bias given in table 2 of \cite{2010ApJ...724..878T}, the halo mass function of \cite{Tinker:2008ff} with the updated parameters given in \cite{2010ApJ...724..878T}, and the halo Compton $y$-parameter measured in \cite{Arnaud:2009tt}.
We also follow the prescription given in \cite{Bolliet:2017lha}, using a single definition of mass ($M_{500,c}$) and interpolate the bias and halo mass function to the correct value.

The SZ powerspectrum is given by the sum of the one- and two-halo terms \cite{Hill:2013baa,2011MNRAS.418.2207T}, which respectively read
\begin{gather}
C_\ell^{1h}  = \int \diff z \frac{\Diff{2} V}{\diff z \diff \Omega} \int \diff M \frac{\diff n}{\diff M} (z, M) |\tilde{y}_\ell(z,M)|^2,
\label{eq:Clyy1h}
\\
C_\ell^{2h}  = \int \diff z \frac{\Diff{2} V}{\diff z \diff \Omega} D_+^2(z) P_m(k) \bigg[\int \diff M \frac{\diff n}{\diff M} (z, M)  b_H(z, M) \tilde{y}_\ell(z,M) \bigg]^2\bigg|_{k=\big(\frac{\ell + 1/2}{\chi(z)}\big)}\; .
\label{eq:Clyy2h}
\end{gather}
Here $P_m(k)$ is the linear matter powerspectrum, $D_+(z)$ is the growth factor, $\Diff{2} V/ \diff z \diff \Omega = c\chi^2(z)/H(z) $ is the comoving volume element per steradians and $\tilde{y}_\ell(z,M)$ is the 2D Fourier transform of the projected $y$-parameter image of the halo
\begin{equation}
\tilde{y}_\ell(z,M) = \frac{4\pi r_{s,y}}{\ell_{s,y}^2} \int \diff x x^2 j_0\left(\frac{k x}{\ell_s}\right) y_{3D}(z,M,x),
\end{equation}
$r_{s,y}$ is the typical scale radius of the $y$-image of the halo and $\ell_{s,y} = a(z) \chi(z)/r_{s,y} $.
We refer to the appendix of \cite{Hill:2013baa} for a clear derivation of these two formulae.
The SZ effect cross correlates with $T$ through the late ISW effect and also correlates with $E$.
The latter effect is given by the fact that after reionization the universe remains ionized and therefore free electrons generate polarization anisotropies even at very low redshift.
SZ is generated in the same epoch on similar scales, so this gives rise to a non-vanishing $y$-$E$.
This is expected to be a very small effect at low redshifts, but we consider it as it might still constitute a sizable bias if the primordial signal is small too.

In our numerical evaluation, we will use the full temperature (polarization) transfer functions $\mathcal{T}^{T,(E)}_\ell (k)$, extracted from \texttt{CLASS} \cite{Lesgourgues:2011re}.
Defining real space temperature transfer functions \cite{Komatsu:2003fd, Liguori:2003mb} as
\begin{equation}
\mathcal{T}^T_\ell (\chi) = \frac{2}{\pi}\int \diff k \, k^2 \mathcal{T}^T_\ell (k) j_\ell(\chi k) \; ,
\end{equation}
and using the Poisson equation, 
$\delta (\vec{k},z)= \frac{3}{5}(\Omega_{M}H_0^2)^{-1} \frac{D(z)}{a(z)} \mathcal{T}_m (k) k^2 \phi(\vec{k}, z=0)$, to express the overdensity contrast as a function of the gravitational potential we can write the temperature-SZ cross correlation:
\begin{equation}
\begin{split}
C_\ell^{\text{SZ-}T} =&\int \frac{c \diff z}{H(z)} \frac{3}{5}\frac{c^2 k^2 \chi^2(z) \mathcal{T}_m (k)}{\Omega_m H_0^2}D_+(z)\int \diff M \frac{\diff n}{\diff M}(z,M) \tilde{y}_\ell\big(z,M) b_H(z,M)\times
\\
&
\times\mathcal{T}^T_\ell (\chi(z))
P_\zeta (k)\bigg|_{k=\big(\frac{\ell + 1/2}{\chi(z)}\big)}.
\end{split}
\label{eq:SZT}
\end{equation}
Notice that one can write the same quantity for polarization just replacing $T\rightarrow E$ in the last two equations.

For an explicit numerical evaluation of these auto- and cross-correlations, we worked in Limber approximation.
This allows removing one of the $5$ nested integrals, making the computation  numerically feasible.
Even though, practically, all the SZ signal comes from $z<4$ \cite{Hill:2013baa}, since $E$ is sourced at reionization we extend all the redshift integrations to $z$ well above the time of reionization. As expected, the contributions from $z>4$ are negligible.
Unless otherwise stated, we choose $z>0.005$ as lower integration limit; this ensures that the redshift integrals do not get contributions from unphysical $z = 0$ objects.
We integrate over the masses $10^{10}M_\odot h^{-1} < M < 10^{16}M_\odot h^{-1}$.

\subsection{Bias, variance, and Fisher results}
We now have all the ingredients to calculate numerically the Fisher matrix for $T$ and $E$ (separately), Eq.~(\ref{eq:FishT}), and for the joint analysis of $T$ and $E$, Eq.~(\ref{eq:FishTE}).

All the results that we will present are obtained marginalizing over $\alpha$, that is the amplitude of the SZ-$T$ and SZ-$E$ cross correlations.
However it is worth noticing that marginalizing over $\alpha$ do not degrade the constraints on $f'$ by much.
To quantify this statement we calculate the correlation between the two parameters.
Those are
\begin{equation}
\frac{\Big[\sum_\ell \frac{\partial C_\ell^{y T}}{\partial f'} \frac{\partial C_\ell^{\text{SZ} T}}{\partial \alpha}\Big]}{\sqrt{\Big[\sum_\ell \Big(\frac{\partial C_\ell^{y T}}{\partial f'} \Big)^2\Big] \Big[\sum_\ell \Big(\frac{\partial C_\ell^{\text{SZ} T}}{\partial \alpha}\Big)^2\Big]}}= 0.28,
\quad 
\frac{\Big[\sum_\ell \frac{\partial C_\ell^{y E}}{\partial f'} \frac{\partial C_\ell^{\text{SZ} E}}{\partial \alpha}\Big]}{\sqrt{\Big[\sum_\ell \Big(\frac{\partial C_\ell^{y E}}{\partial f'} \Big)^2\Big] \Big[\sum_\ell \Big(\frac{\partial C_\ell^{\text{SZ} E}}{\partial \alpha}\Big)^2\Big]}}= 0.53.
\label{eq:SZT-yTcorr}
\end{equation}
In the limit of 0 correlation, omitting $f_\text{sky}$ and considering only the temperature anisotropies, the signal-to-noise ratio (SNR) for the parameter $f'$ would have been 
\begin{equation}
\text{SNR}(f') \approx \sqrt{\sum_\ell \frac{ (C_\ell^{yT})^2}{C_\ell^{TT} (C_\ell^\text{SZSZ}+C_\ell^{yyN})}}\, .
\end{equation}
We are not using this approximate formula in the analysis, but it is useful to clarify the methodology.
Since we do not have any control over the signal, the only way to increase the SNR is to try to minimize the noise.
The temperature anisotropies measurements are already dominated by cosmic variance up to very high multipoles.
The variance of the $y$-distortions is instead composed by two contributions: the cosmic variance term and the instrumental noise.
We consider different experimental setup to understand what is the sensitivity needed by future experiments to reach the lower limit imposed by cosmic variance.
The noise term is $C_\ell^{yy,N} = 4\pi \times N \times e^{\ell^2/\ell_\text{beam}^2}$;
where $(N, \ell_\text{beam})$ is 
$(0, -)$ for the CVL survey,
$(4 \times10^{-18}, 84)$ for PIXIE, $(4 \times10^{-20}, 100)$ for the PRISM spectrometer (PRISM spec) and $(4 \times10^{-20}, 4000)$ for the PRISM imager, which could be easily calibrated using the on-board spectrometer but is not bounded by low angular resolution \cite{Ganc:2012ae, Emami:2015xqa, Ravenni:2017lgw}.
Since we are interested in differential measurements of $y$ distortions, the LiteBIRD satellite can also be employed~\cite{Matsumura:2013aja}.
Differential measurements of small signals still rely on very precise inter-channel calibration of the instrument that will probably be challenging if not unfeasible without a calibrator on board \cite{Ganc:2012ae}.
This point is even more crucial if we consider that component separation will be required to discern the foregrounds.
As shown in \cite{Remazeilles:2018kqd}, the calibration relative errors must be of order 0.01\% in order to recover the correct $\mu$-$T$ cross correlation.
However, assessing correctly the effect of inter-channel calibration error is beyond the scope of this paper.
Instead, we use two different noise levels for LiteBIRD:
an optimistic estimate (Opt) assuming perfect inter-channel calibration $(2 \times10^{-20}, 200)$ and a more conservative (Cons) estimate discussed below.
From \cite{Remazeilles:2018kqd}, we can assess that the ratio of the PIXIE and LiteBIRD noise contribution (neglecting all foregrounds) to $C_\ell^{\mu\mu}$ is $C_\ell^{\mu\mu\text{, Noise, LiteBIRD}}/C_\ell^{\mu\mu\text{, Noise, PIXIE}}=0.06$.
Here we assume that the noise contribution to $C_\ell^{yy}$ follows the same scaling for the two experiments.
Therefore, the parameters we use for LiteBIRD - Cons are $(0.24 \times10^{-18}, 200)$, though we acknowledge that this is a very coarse, order-of-magnitude, estimate.
However, any finer estimate would be beyond the intrinsic uncertainties of a Fisher matrix analysis.
Moreover, our result will not be very sensitive to this parameter because it does not affect the other major contribution to the noise: the SZ effect.

Since the dominant contribution to the SZ powerspectrum is given by big galaxy clusters at low redshift \cite{Aghanim:2015eva}, masking resolved clusters is a viable way to reduce the overall noise \cite{Hill:2013baa, Creque-Sarbinowski:2016wue, Ravenni:2017lgw}.
Following the procedure we outlined in Ref. \cite{Ravenni:2017lgw}, we consider the use of two different sky masks based on two different full sky surveys that will produce maps of galaxy clusters.
We will present the forecasts based on the PIXIE-like and LiteBIRD satellite both for the unmasked case and with a mask based on eROSITA \cite{Merloni:2012uf} expected performance.
We will assume to mask all the clusters with mass greater than $2\times 10^{14}M_\odot/h$ at $z<0.15$.
The forecast for PRISM and the CVL instrument will similarly involve the unmasked case, and a mask based on the expected PRISM cluster catalogue. 
We will assume to mask all the clusters with mass greater than $10^{13}M_\odot h^{-1}$.
We chose those regions of the mass-redshift plane because the eROSITA and PRISM catalogues are expected to be complete in those respective areas.
We summarize the mask boundaries in table \ref{tab:masks}.

In our analysis we account for the effect of masking the galactic plane and resolved cluster according with eq. (\ref{eq:Foperative}).
For PIXIE, LiteBIRD, and PRISM, we assume to use an $f_\text{sky}$ similar to the one used by \textit{Planck} to calculate the SZ power spectrum \cite{Aghanim:2015eva}, namely $f_\text{sky} = 50\%$ to remove the galactic plane.
To account for the reduced sky area after applying the eROSITA and PRISM mask, we reduce $f_\text{sky}$ to 35\%. According with \cite{Ravenni:2017lgw}, this is a conservative estimate for both cluster masks.
In the case of the CVL survey we show the results assuming $f_\text{sky} = 100\%$, since they have to be intended as upper limits that might in principle be reached with other techniques that we are not investigating here.

\begin{table}
\centering
\begin{tabular}{|cc|}
\hline
Survey		&	Clusters masked\\
\hline
Unmasked            & None   \\
eROSITA             & $M>2\times 10^{14}M_\odot/h$ for $z<0.15$    \\
PRISM                & $M>10^{13}M_\odot h^{-1}$\\
\hline
\end{tabular}
\caption{We assume to use each survey to mask all clusters with the given characteristics.}
\label{tab:masks}
\end{table}

\begin{figure}
\centering
\includegraphics[width=0.9\textwidth]{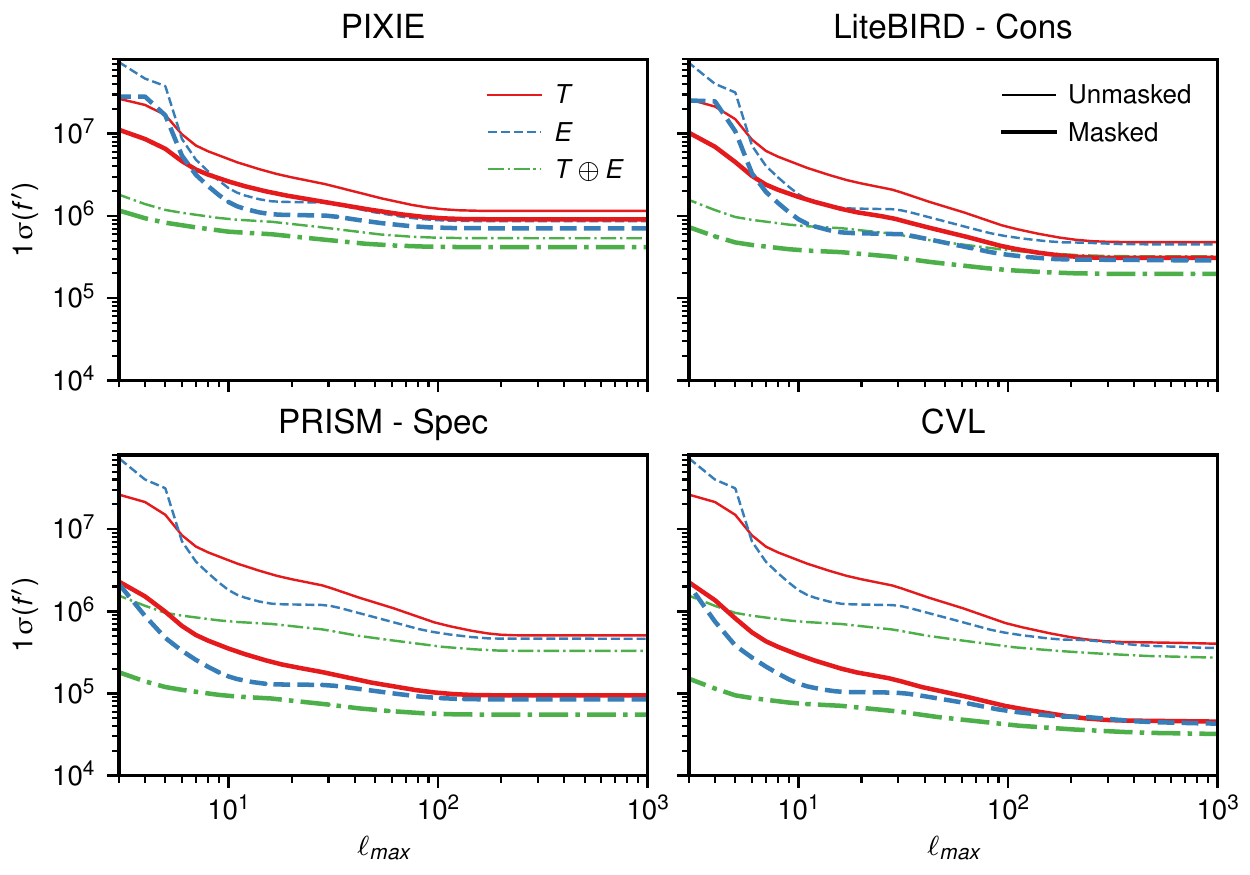}
\caption{Forecasted 1 $\sigma$ error on $f'$. We compare the different experimental setup, and evaluate their performance both masking and not-masking the most massive clusters in the $y$-distortions map. We consider the cross correlation with temperature and polarization anisotropies, taken one at a time and analyzed jointly.
We consider the mask based on eROSITA for PIXIE and LiteBIRD, and the one based on PRISM for PRISM itself and the CVL experiment.
}
\label{fig:Results}
\end{figure}

In figure \ref{fig:Results} we show the 1$\sigma$ forecasted error bars on $f'$ as a function of the max $\ell$ considered.
The same results are summarized in table \ref{tab:results}.
There we assume the use of all the multipoles available in each experimental configuration, i.e., to relate these results with figure \ref{fig:Results} one has to take $\ell_\text{max}\gg \ell_\text{beam}$.
We notice that for the linearity of $C_\ell^{yT(E)}$ with respect to $f'$, the forecasted error bars are independent from the chosen $f'$ fiducial value.
As it is clear from figure \ref{fig:Results}, none of the experimental setups is bounded by raw detector sensitivity.
The two main limiting factors are the SZ powerspectrum --- more and more aggressive masks give better results --- and the survey beam.
Since the primordial signal increases as $\ell$ increases, it is convenient to exploit the higher multipoles to extract some degree of information.
Trying to exploit relatively high multipoles ($\ell \approx $ few hundreds) is beneficial also for an instrument with higher detector sensitivity than PIXIE.
The PRISM imager has much higher angular resolution than the on-board spectrometer.
It is therefore possible to envision its use to make differential measurements of the spectral distortion anisotropies up to small angular scales.
In this case uncertainties in the inter-channel calibration would not be a problem, since the spectrometer can provide a reference spectrum.
If we consider the use of the imager to achieve higher angular resolution, PRISM will basically reach the limit set by cosmic variance.

\begin{table}
\centering
\begin{tabular}{|c|ccc|ccc|}
\hline
	&\multicolumn{3}{|c|}{Unmasked}		&	\multicolumn{3}{c|}{eROSITA mask}\\

				&	$T$					& $E$  		& $T\oplus E$	&	$T$		& $E$	& $T\oplus E$ \\
\hline
PIXIE			&	$11\times 10^5$ 	&	$8.6\times 10^5$	&	$5.3\times 10^5$		&	$9.1\times 10^5$	&	$7.1\times 10^5$	&	$4.2\times 10^5$	\\
LiteBIRD - Opt	&	$4.3\times 10^5$	&	$4.0\times 10^5$	&	$2.9\times 10^5$		&	$2.5\times 10^5$	&	$2.4\times 10^5$	&	$1.7\times 10^5$ \\
LiteBIRD - Cons	&	$4.8\times 10^5$	&	$4.5\times 10^5$	&	$3.2\times 10^5$		&	$3.1\times 10^5$	&	$2.9\times 10^5$	&	$2.0\times 10^5$\\
\hline
	&\multicolumn{3}{|c|}{Unmasked}		&	\multicolumn{3}{c|}{PRISM mask}\\
\hline
PRISM - spec		&	$5.1\times 10^5$	&	$4.6\times 10^5$	&	$3.3\times 10^5$		&	$9.5\times 10^4$	&	$8.5\times 10^4$	&	$5.5\times 10^4$\\
PRISM - imager		&	$4.1\times 10^5$	&	$3.6\times 10^5$	&	$2.8\times 10^5$		&	$7.0\times 10^4$	&	$6.7\times 10^4$	&	$4.7\times 10^4$\\
CVL					&	$2.8\times 10^5$	&	$2.5\times 10^5$	&	$1.9\times 10^5$		&	$3.0\times 10^4$	&	$2.9\times 10^4$	&	$2.2\times 10^4$\\
\hline
\end{tabular}
\caption{1$\sigma$ forecasted error bars on $f'$. $T$, $E$, and $T\oplus E$ indicate respectively the forecast using temperature, polarization, and both.
In all the forecasts we marginalize over the amplitude of $C_\ell^{\text{SZ}T}$ and $C_\ell^{\text{SZ}E}$.
}
\label{tab:results}
\end{table}

\section{Conclusions}
\label{sec:Conclusions}

In this paper, we provided a new framework to calculate the linear fluctuation of the spectral $y$-distortions.
It was shown that a solution to the Boltzmann equation for the Compton scattering can be constructed from 4 parameters including the temperature perturbation and the spectral $y$-distortion.
Then we derived the evolution equation for the $y$-distortion, which is sensitive to the baryon isocurvature perturbations.
This implies that it can resolve the degeneracy between baryon isocurvature perturbations and the CDM ones in contrast to the standard linear perturbations such as the temperature perturbations and the polarizations. 
We numerically estimated the transfer function of the $y$-distortions based on \texttt{CLASS} and computed the auto and the cross correlations with the temperature perturbations and the polarization $E$ modes.
The resulting $C_{\ell}$'s completely resolve the degeneracy between baryon isocurvature perturbations and the CDM isocurvature perturbations as we expected, and we found that only the correlated CIPs contribute to them.
The auto correlation of the spectral $y$-distortions is strongly contaminated by SZ powerspectrum and therefore we could not get any information on the $f_{\rm bi}$ parameter alone from it.
Note that linear $y$ anisotropies are not contaminated by lensing effect in contrast to the previous methods based on the nonlinear modulation of the CMB anisotropy.
Then we produced a forecast for the upper bounds on correlated CIPs for different future observational projects.
Even in the absence of foregrounds, none of the surveys we consider will be able to set stringent constraints on CIPs.
For instance, $f'< 2 \times 10^{5}$ at 68\% C.L. is obtained for LiteBIRD, while $f'<5 \times 10^{4}$ for PRISM, and $f'<2\times 10^{4}$ for a cosmic variance limited survey.
As we have shown, the fundamental limit is set by the noise contribution due to the SZ powerspectrum.
However it is important to remember that
our method can resolve the degeneracy between the correlated and uncorrelated CIPs.
This implies that we can in principle distinguish the correlated and uncorrelated CIPs by combining our analysis with the other methods discussed before.
Our method would be more useful if more powerful techniques were developed to remove the SZ-induced noise term in the future, though we have to thoroughly assess not only the SZ effect but also all foregrounds in this case.

We do not focus on the specific models that produce the CIPs in this paper.
As discussed in Ref.~\cite{He:2015msa}, models based on the curvaton only produce CIPs as big as $f_{\rm bi}<16$, while the spontaneous baryogenesis in Ref.~\cite{DeSimone:2016ofp} does not generate the correlated CIPs.
Therefore, the known scenarios do not expect huge $f'$, which we may observe in our method; nevertheless, it would be interesting as the author in Ref.~\cite{Valiviita:2017fbx} reported the possibility of sizable CIPs.
In other words, highly nontrivial early universe physics would be suggested if significant CIP were detected.
 
In our calculations we dropped $\mathcal O(\delta^{3})$ terms for simplicity, but these may have comparable contributions to the spectral distortions for the Gaussian adiabatic perturbations.
Therefore, cumbersome cubic order analysis would be required if the sizable linear $y$-distortions are really detected in the future.
We did not discuss the $\kappa$ distortion and the $u$ distortion, which is introduced for the first time in this paper, because they do not explicitly depend on the baryon isocurvature perturbations.
Still it would be interesting to consider the anisotropy of these new spectral distortions in different contexts.
This would be investigated in our future works.

\acknowledgments 

We would like to thank Jens Chluba for several crucial comments and discussions.
We would like to thank Masahide Yamaguchi, Takeshi Kobayashi, Daniel Grin, and Luke Hart for helpful comments.
K.I. is supported by World Premier International Research Center Initiative (WPI Initiative), MEXT, Japan, 
Advanced Leading Graduate Course for Photon Science, 
and JSPS Research Fellowship for Young Scientists.
A.O. is supported by JSPS Overseas Research Fellowships.
A.R. thanks the University of Manchester and the Jodrell Bank Centre for Astrophysics for hospitality.


\bibliographystyle{unsrt}

\end{document}